%% file: j_phys_a_2006.tex
\documentclass[12pt]{iopart}
\usepackage[dvipdfm]{graphicx}
\begin{document}

\title[A new stochastic cellular automaton model on traffic ...]
{A new stochastic cellular automaton model on traffic flow
and its jamming phase transition}

\author{Satoshi Sakai$^1$, Katsuhiro Nishinari$^2$, and Shinji Iida$^1$}

\address{$^1$Department of Applied Mathematics and Informatics, Ryukoku University, Shiga 520-2194, Japan}
\address{$^2$Department of Aeronautics and Astronautics, Tokyo University, Tokyo 113-8656, Japan}
\begin{abstract}
A general stochastic traffic cellular automaton (CA) model,
which includes slow-to-start effect and driver's perspective,
is proposed in this paper.
It is shown that this model includes well known traffic CA models
such as Nagel-Schreckenberg model, Quick-Start model, and Slow-to-Start model
as specific cases. Fundamental diagrams of this new model clearly
show metastable states around the critical density even
when stochastic effect is present.
We also obtain analytic expressions of
the phase transition curve in phase diagrams by using approximate
flow-density relations at boundaries.
These phase transition curves are in excellent agreement
with numerical results.
\end{abstract}

\maketitle

\section{Introduction}
\label{sec:Traffic_CA}
A traffic jam is one of the most serious issues in modern society.
In Japan, the amount of financial loss due to traffic jam approximates
$12$ thousand billion yen per year according to
Road Bureau, Ministry of Land, Infrastructure and Transport.

Recently, investigations toward understanding traffic jam formation
are done not only by engineers but also actively
by physicists \cite{review_traffic}.
The typical examples are: developing realistic mathematical traffic models
which give (with help of numerical simulations)
various phenomena that reproducing empirical traffic flow
\cite{NS, SlS, Barlovic, FI, QS, NFS, SOV_Rapid};
fundamental studies of traffic models such as obtaining exact solutions,
clarifying the structure of flow-density diagrams
or phase diagrams \cite{GoE_SS, ASEP, DLG_ASEP_PD, Cecile};
analysis of traffic flow with bottleneck \cite{bottleneck_IF, OV_YS};
and studies of traffic flow in various
road types \cite{CrossRoad, Jiang, Huang}.

We can classify microscopic traffic model into two kinds;
optimal velocity models and cellular automaton (CA) models.
A merit of using a CA model is that,
owing to its discreteness, it can be expressed in relatively simple rules
even in the case of complex road geometry.
Thus numerical simulations can be effectively performed and
various structures of roads with multiple lanes can be easily incorporated
into numerical simulations.
However, the study of traffic CA model has relatively short history and
we have not yet obtained the ``best'' traffic CA model which should be
both realistic as well as simple.
There are lots of traffic CA models proposed so far.
For example Rule-$184$ \cite{ECA},
which was originally presented by Wolfram as a part of Elementary CA,
is the simplest traffic model.
Fukui-Ishibashi (FI) model \cite{FI} takes into account
high speed effect of vehicles,
Nagel-Schreckenberg (NS) model \cite{NS} deals with random braking effect,
Quick-Start (QS) model \cite{QS} with driver's anticipation effect and
Slow-to-Start (SlS) model \cite{SlS} with inertia effect of cars.
Asymmetric Simple Exclusion Process (ASEP) \cite{ASEP},
which is a simple case of the NS model,
has been often used to describe general nonequilibrium systems
in low dimensions.
An extension of the NS model, called VDR model, is considered
by taking into account a kind of slow-to-start effect \cite{review_traffic}.
Recently Kerner et al proposed an elaborated CA model by taking into account
the synchronized distance between cars \cite{Kerner}.
Each of these models reproduces a part of features of  empirical traffic flow.

\begin{figure}[tbph]
\begin{center}
\includegraphics[width=0.7\linewidth]{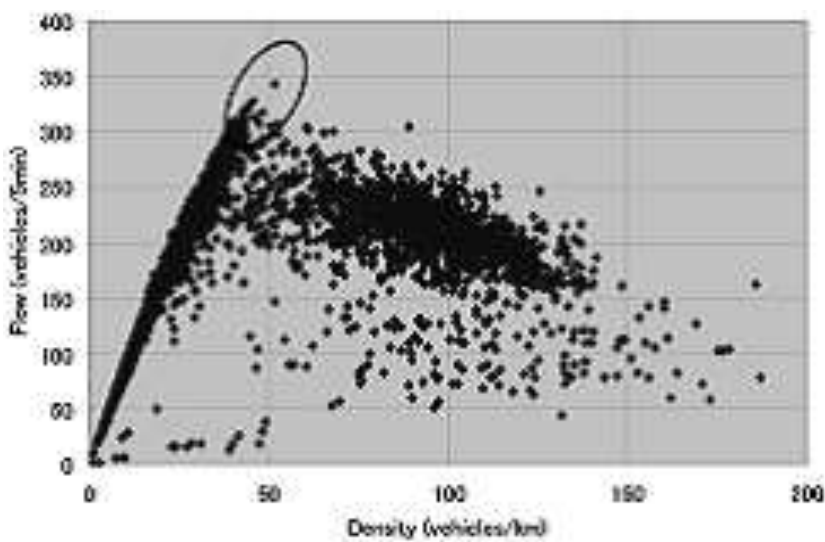}
\end{center}
\caption{A fundamental diagram is plotted with use of empirical data.
These data were collected at the Tomei expressway in Japan
by Japan Highway Public Corporation.
The road consists of two-lanes.
Each points in the diagram corresponds to an average over a time interval
of five minutes.
The density is calculated from the flow divided by the average velocity.
We can see metastable points which are inside the circle
at the highest flux region of free flow.}
\label{fig:ActualData}
\end{figure}

A fundamental diagram is usually used to examine
whether a model is practical or not by comparing empirical data
(see figure \ref{fig:ActualData}) and simulation results.
A fundamental diagram is called a flow-density graph in other words.
And it chiefly consists of three parts: free-flow line,
jamming line, and metastable branches.
On the free-flow line, flow increases with density,
while on the jamming line flow decreases with density.
The critical density divides free-flow line from jamming line.
Around the critical density, close to the maximal flow,
there appear some metastable branches.
In general, the metastable branches are unstable states
where cars can run like free-flow state
even if the density surpasses the critical density.
Characteristics of individual fundamental diagrams for above CA models are
as follows:
the fundamental diagram of Rule-$184$ has an isosceles triangular shape.
The critical density in FI model, that divides the whole region into
free flow phase and jamming phase, is lower than that of Rule-184,
and the maximal flow reaches a higher value due to the high speed effect.
The critical density and maximal flow in QS model are higher than
those of Rule-$184$, because more than one cars can move simultaneously.
The maximal flow of ASEP or NS model is lower than those of Rule-$184$ or
FI model because of random braking effect.
In the case of SlS model, a fundamental diagram is significantly different from
these models. Its fundamental diagram shows metastable branches
near the critical density \cite{SlS, Barlovic}.
Because metastable branches are clearly observed
in practical fundamental diagram (see figure \ref{fig:ActualData}),
we think realistic traffic models should reproduce such branches.
We note that the study of these metastable branches in the high flow region
is very important for a plan of ITS (Intelligent Transport Systems).
The states in these branches are dangerous, because of the lack of headway
between vehicles.
On the other hand, they have the highest transportation efficiency
because each vehicle's speed is very high regardless of the short headway.

In this paper, we propose a general stochastic traffic CA model
which includes, as special cases, well known models such as NS model,
QS model, and SlS model.
The paper is organized as follows.
We first define our model in sec. $2$.
Next, fundamental diagrams, flow-$\alpha$-$\beta$ diagrams and
phase diagrams of this new model will be presented in sec. $3$ and $4$.
Finally we show analytical expressions of
phase transition curves in phase diagrams in sec. $5$.

\section{A new stochastic CA model}
In 2004, a new deterministic traffic model
which includes both of slow-to-start effects
and driver's perspective (anticipation), was presented by
Nishinari, Fukui, and Schadschneider \cite{NFS}.
We call this model Nishinari-Fukui-Schadschneider (NFS) model.
In this paper we further extend NFS model by incorporating stochastic effects.

First, the updating rules of NFS model are written as
\begin{eqnarray}
\label{eq:NFS1}
v_{i}^{(1)} = \min \{ V_{\max},\; v_{i}^{(0)}+1 \}\\
\label{eq:NFS2}
v_{i}^{(2)} = \min \{ v_{i}^{(1)},\; x_{i+S}^{t-1} - x_{i}^{t-1} - S \}\\
\label{eq:NFS3}
v_{i}^{(3)} = \min \{ v_{i}^{(2)},\; x_{i+S}^{t} - x_{i}^{t} - S \}\\
\label{eq:NFS4}
v_{i}^{(4)} = \min \{ v_{i}^{(3)},\; x_{i+1}^{t} - x_{i}^{t} - 1 + v_{i+1}^{(3)} \}\\
\label{eq:NFS5}
x_{i}^{t+1} = x_{i}^{t} + v_{i}^{(4)}
\end{eqnarray}
in Lagrange representation.
In these rules, $x_{i}^{t}$ is a Lagrange variable
that denotes the position of the $i$th car
at time $t$. $v_{i}^{(0)}$ is velocity $x_{i}^{t} - x_{i}^{t-1}$,
and the parameter $S$ represents the interaction horizon of drivers
and is called an anticipation parameter.
And we repeat and apply this rule again
as $v_{i}^{(0)} \longleftarrow v_{i}^{(4)}$ in the next time.
In this paper, we use a parallel update scheme where those rules are applied
to all cars simultaneously.
Rule \mbox{(\ref{eq:NFS1})} means acceleration and the maximum velocity
is $V_{\max}$,
\mbox{(\ref{eq:NFS2})} realizes slow-to-start effect,
\mbox{(\ref{eq:NFS3})} means deceleration due to other cars,
\mbox{(\ref{eq:NFS4})} guarantees avoidance of collision.
Cars are moved according to \mbox{(\ref{eq:NFS5})}.
The characteristic of NFS model is the occurrence of metastable branch in
a fundamental diagram because of the slow-to-start effect.
It should be noted that the slow-to-start rule adopted
in this paper (\ref{eq:NFS2}) is different from
the previously proposed ones \cite{Barlovic}.

\begin{figure}[htbp]
\begin{center}
\input{figure2.tex}
\caption{A reduction map of various traffic CA models.}
\label{fig:S-NFS_allModel}
\end{center}
\end{figure}
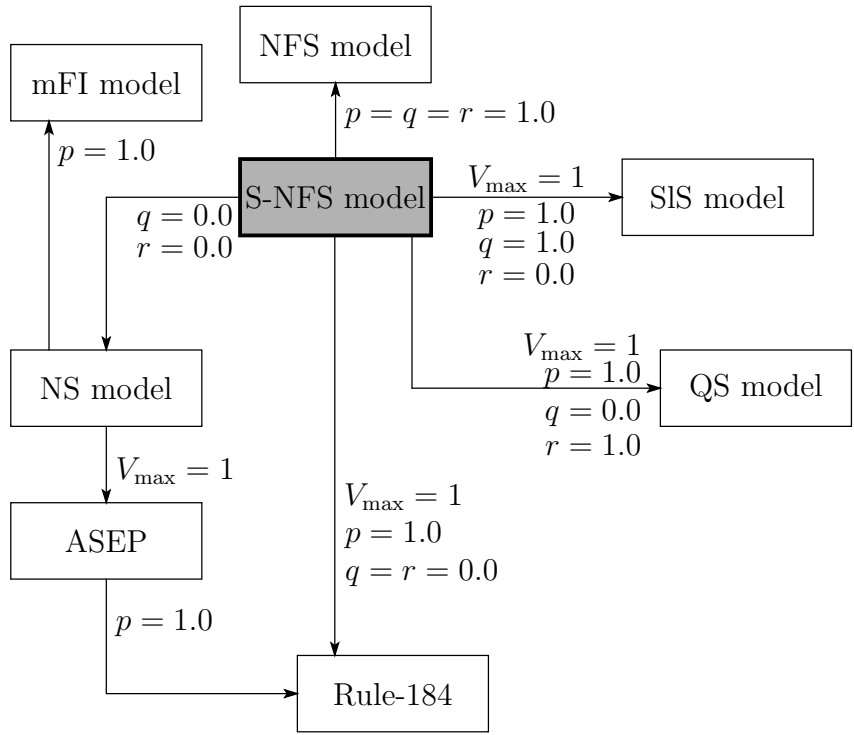

Next, we explain a stochastic extension of NFS model.
The model is described as follows:
\begin{eqnarray}
\label{eq:S-NFS1}
v_{i}^{(1)} = \min \{ V_{\max},\; v_{i}^{(0)}+1 \}\\
\label{eq:S-NFS2}
v_{i}^{(2)} = \min \{ v_{i}^{(1)},\; x_{i+S}^{t-1} - x_{i}^{t-1} - S \} && \mbox{with the probability } q\\
\label{eq:S-NFS3}
v_{i}^{(3)} = \min \{ v_{i}^{(2)},\; x_{i+S}^{t} - x_{i}^{t} - S \}\\
\label{eq:S-NFS4}
v_{i}^{(4)} = \max \{ 0,\; v_{i}^{(3)}-1 \} && \mbox{with the probability } 1-p\\
\label{eq:S-NFS5}
v_{i}^{(5)} = \min \{ v_{i}^{(4)},\; x_{i+1}^{t} - x_{i}^{t} - 1 + v_{i+1}^{(4)} \}\\
\label{eq:S-NFS6}
x_{i}^{t+1} = x_{i}^{t} + v_{i}^{(5)}.
\end{eqnarray}
We define independent three parameters, $p$, $q$, and $r$:
the parameter $p$ controls random-braking effect;
that is, a vehicle decreases its velocity by $1$ with the probability $1-p$.
The parameter $q$ denotes the probability that slow-to-start effect is on;
the rule \mbox{(\ref{eq:S-NFS2})} is effective with the probability $q$.
$r$ denotes the probability of anticipation; $S = 2$ with the probability $r$
and $S = 1$ with $1-r$, hence the average
of $S$ is $\langle S \rangle = 1+r$.
This value of $S$ is used in rules \mbox{(\ref{eq:S-NFS2})} and
\mbox{(\ref{eq:S-NFS3})}.
We note that in order to reproduce empirical fundamental diagram
the value of $S$ should lie between $S=1$ and $S=2$ \cite{NFS}.
Thus we call this new extended model Stochastic(S)-NFS model.
A remarkable point is that the S-NFS model includes most of the other models
when parameters are chosen specifically (see figure \ref{fig:S-NFS_allModel}).
Here, ``modified(m) FI model'' means a
modification of the original FI model
so that the increase of car's velocity is
at most unity per one time step,
because, in the original model, cars can suddenly accelerate,
for example, from velocity $v = 0$ to $v = V_{\max}$,
which seems to be unrealistic.

\section{Fundamental diagram}
\begin{figure}[hbpt]
\begin{center}
\begin{tabular}{ccc}
 \begin{minipage}{0.30\hsize}
  \begin{center}
   \includegraphics[width=\hsize]{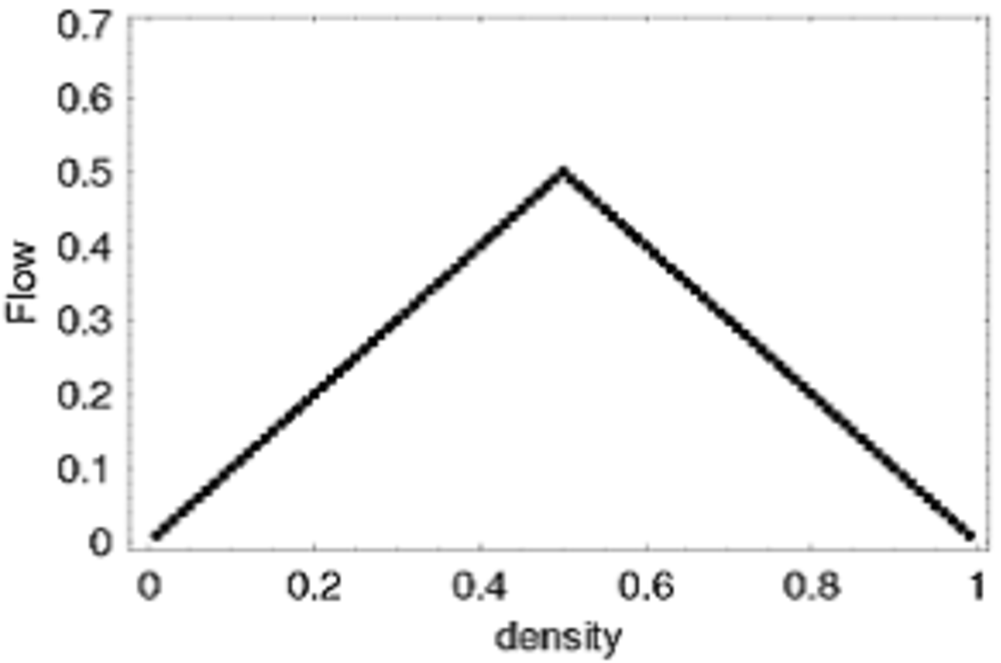}\\
(a)$q = 0.0$, $r = 0.0$\\
  \end{center}
 \end{minipage}
&
 \begin{minipage}{0.30\hsize}
  \begin{center}
   \includegraphics[width=\hsize]{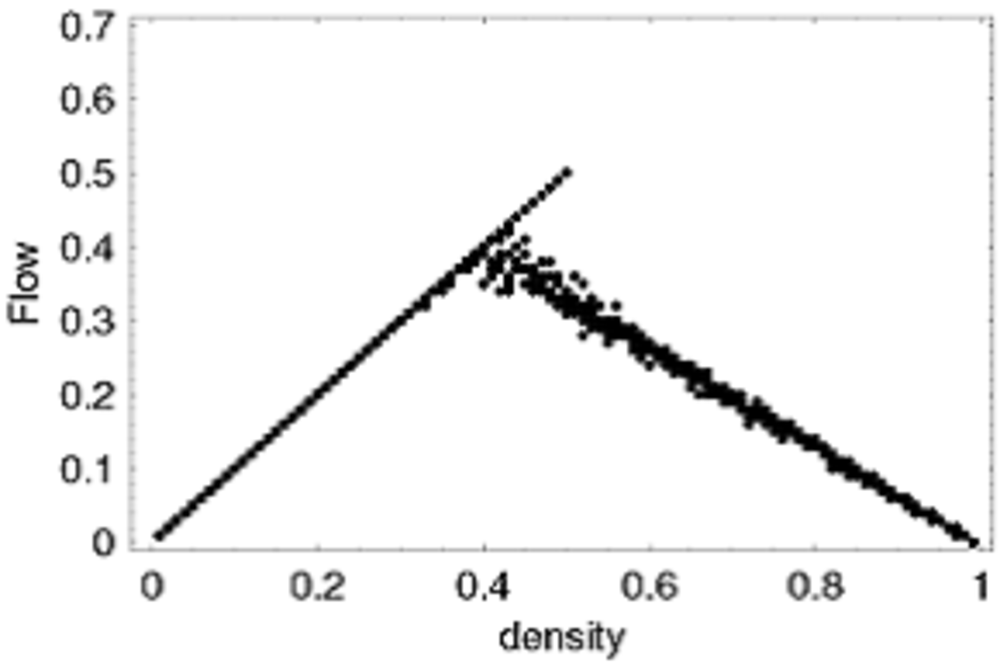}\\
(b)$q = 0.5$, $r = 0.0$\\
  \end{center}
 \end{minipage}
&
 \begin{minipage}{0.30\hsize}
  \begin{center}
   \includegraphics[width=\hsize]{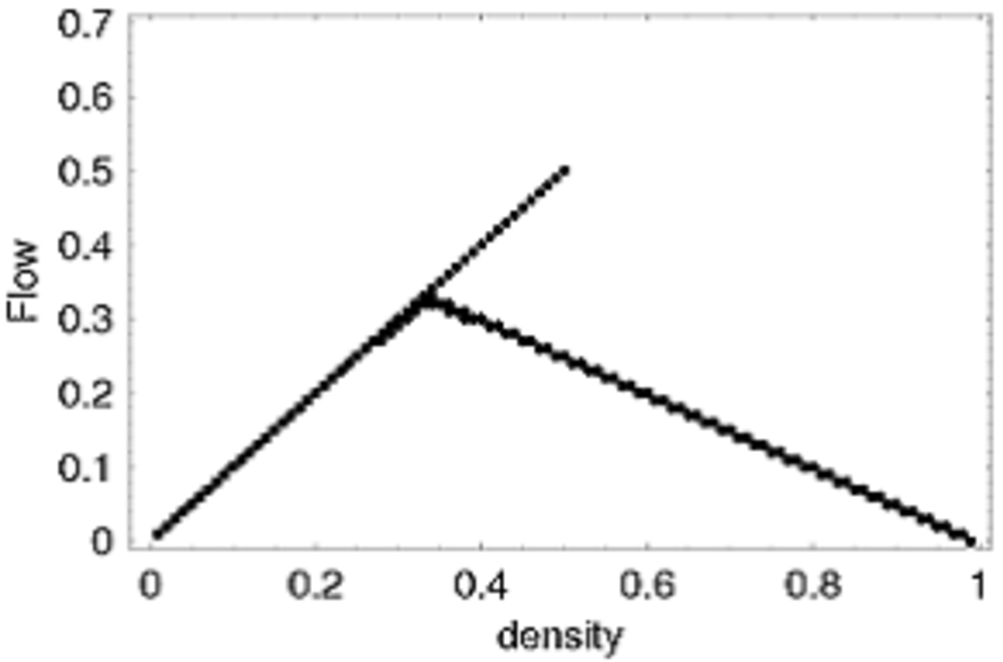}\\
(c)$q = 1.0$, $r = 0.0$\\
  \end{center}
 \end{minipage}\\ \\
 \begin{minipage}{0.30\hsize}
  \begin{center}
   \includegraphics[width=\hsize]{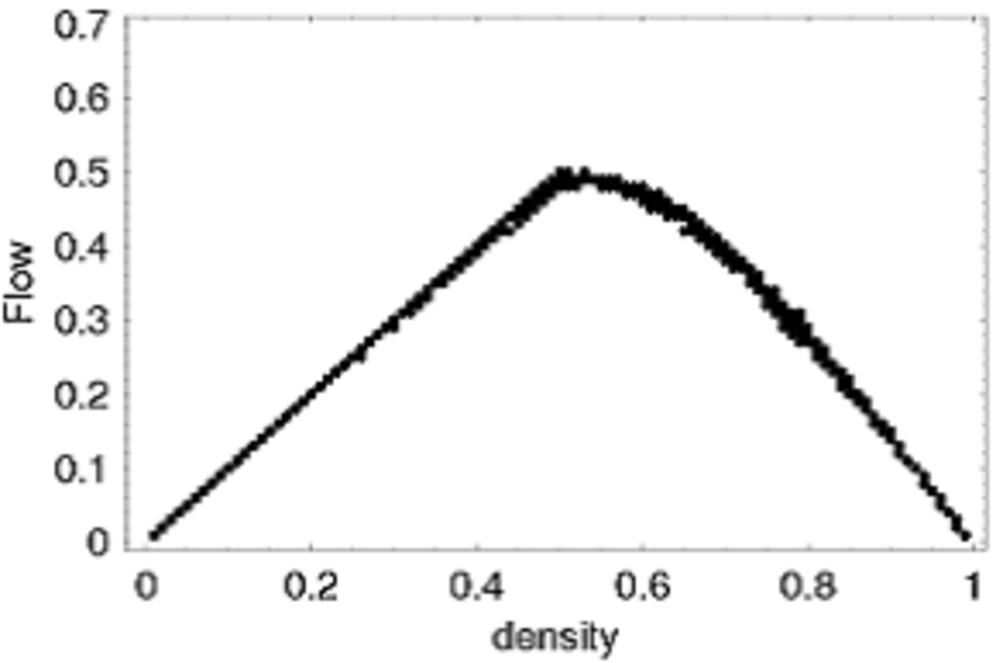}\\
(d)$q = 0.0$, $r = 0.5$\\
  \end{center}
 \end{minipage}
&
 \begin{minipage}{0.30\hsize}
  \begin{center}
   \includegraphics[width=\hsize]{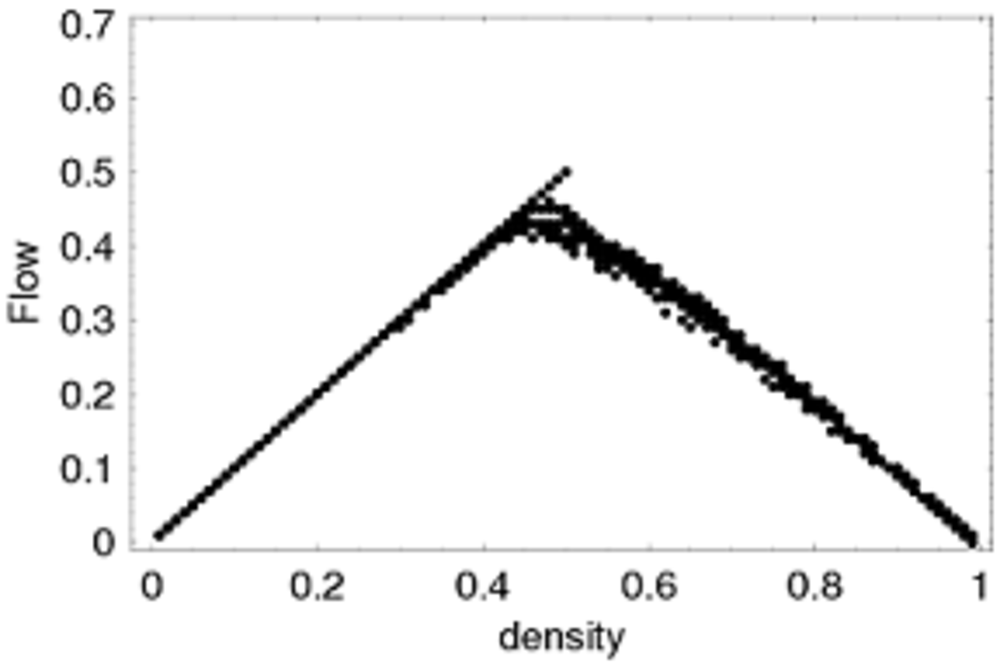}\\
(e)$q = 0.5$, $r = 0.5$\\
  \end{center}
 \end{minipage}
&
 \begin{minipage}{0.30\hsize}
  \begin{center}
   \includegraphics[width=\hsize]{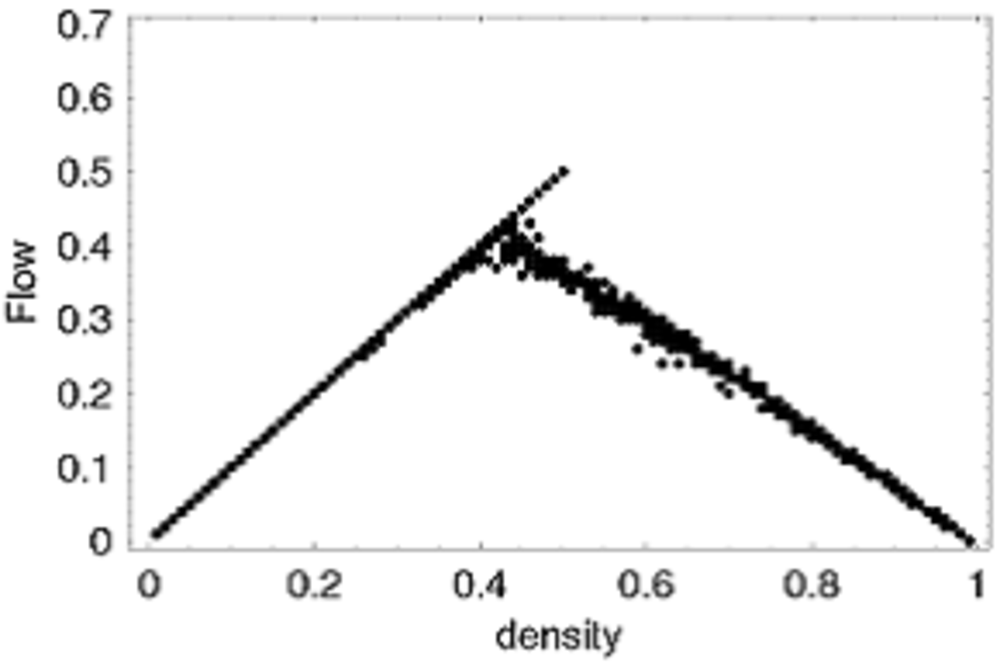}\\
(f)$q = 1.0$, $r = 0.5$\\
  \end{center}
 \end{minipage}\\ \\
 \begin{minipage}{0.30\hsize}
  \begin{center}
   \includegraphics[width=\hsize]{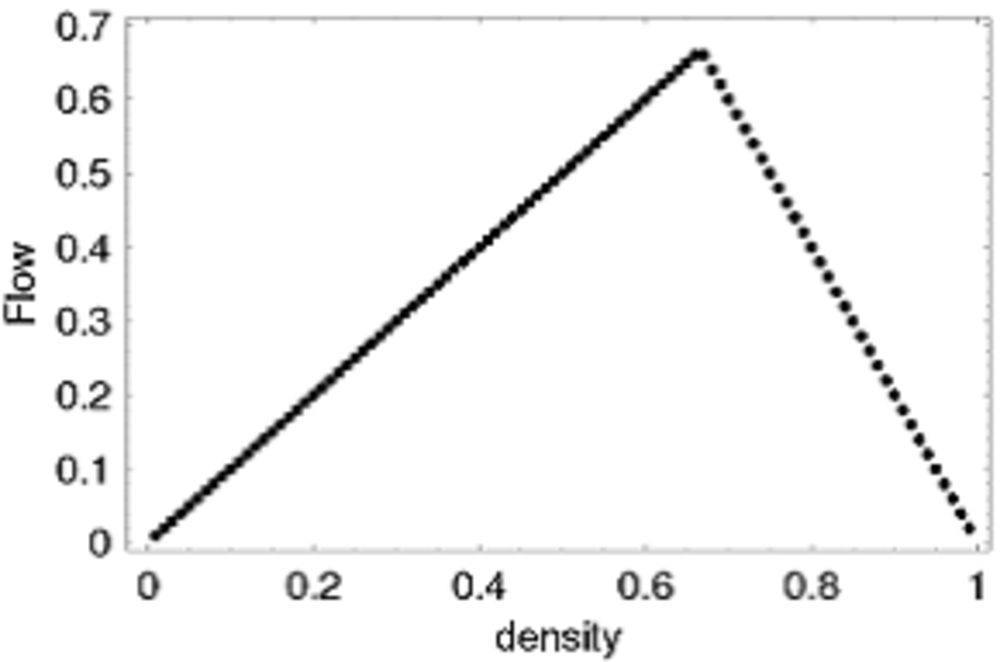}\\
(g)$q = 0.0$, $r = 1.0$\\
  \end{center}
 \end{minipage}
&
 \begin{minipage}{0.30\hsize}
  \begin{center}
   \includegraphics[width=\hsize]{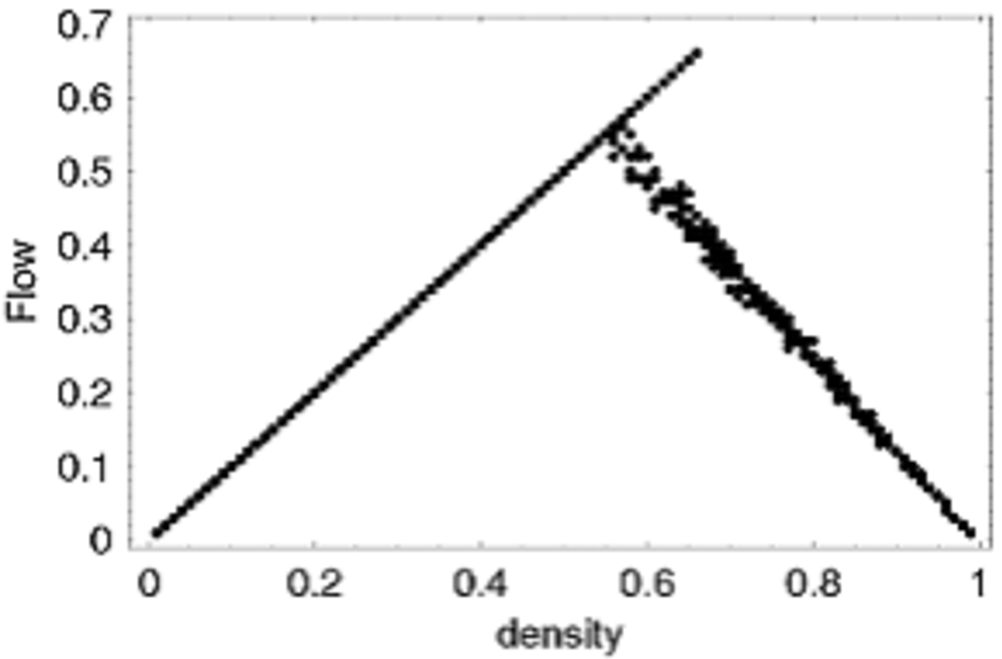}\\
(h)$q = 0.5$, $r = 1.0$\\
  \end{center}
 \end{minipage}
&
 \begin{minipage}{0.30\hsize}
  \begin{center}
   \includegraphics[width=\hsize]{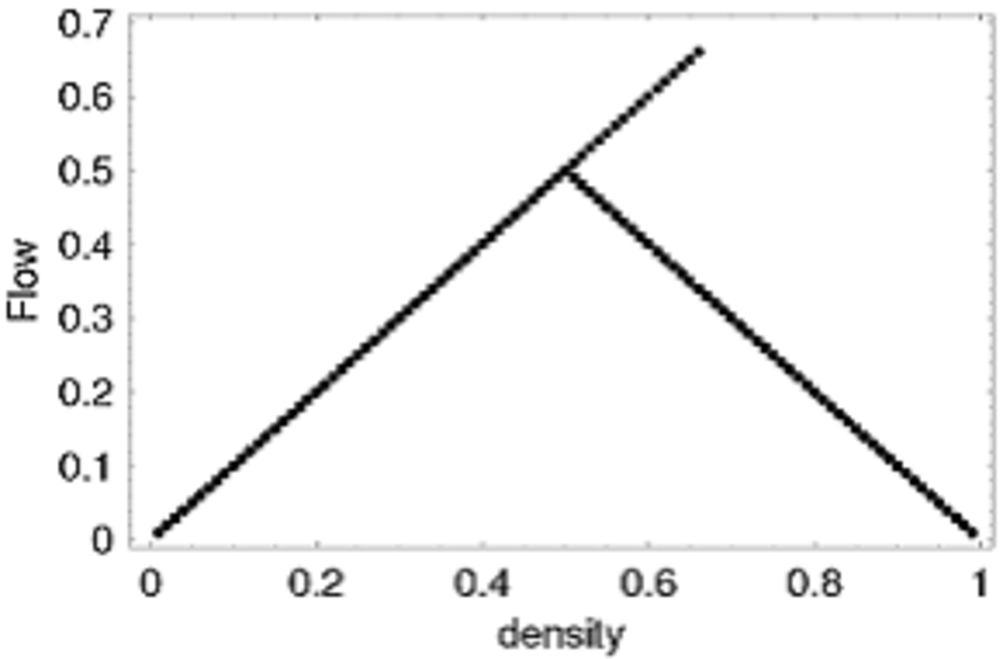}\\
(i)$q = 1.0$, $r = 1.0$\\
  \end{center}
 \end{minipage}
\end{tabular}
\end{center}
\caption{Fundamental diagrams of S-NFS model with $p = 1.0$, 
$L = 100$ and $V_{\max} = 1$.
We repeat simulations for 10 initial 
conditions at each density. 
The flow is averaged over the range from $t = 50$ to $t = 100$.
When $q \neq 0$ and for uniform initial distributions,
metastable branches are clearly seen.}
\label{fig:S-NFS_FD}
\end{figure}

\begin{figure}[hbpt]
\begin{center}
\begin{tabular}{ccc}
 \begin{minipage}{0.30\hsize}
  \begin{center}
   \includegraphics[width=\hsize]{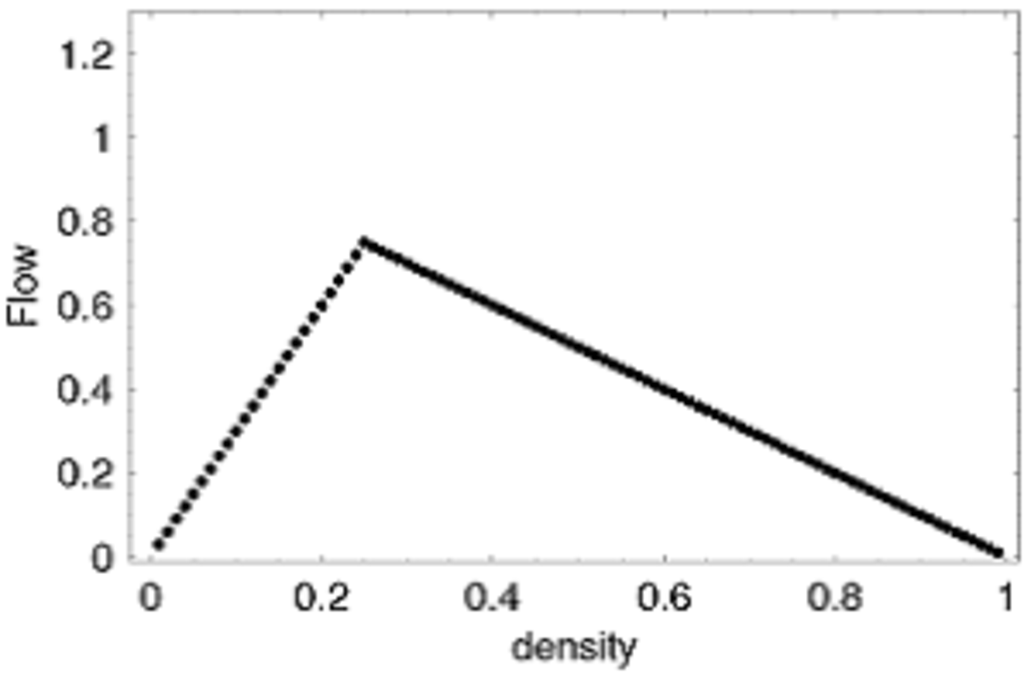}\\
(a)$q = 0.0$, $r = 0.0$\\
  \end{center}
 \end{minipage}
&
 \begin{minipage}{0.30\hsize}
  \begin{center}
   \includegraphics[width=\hsize]{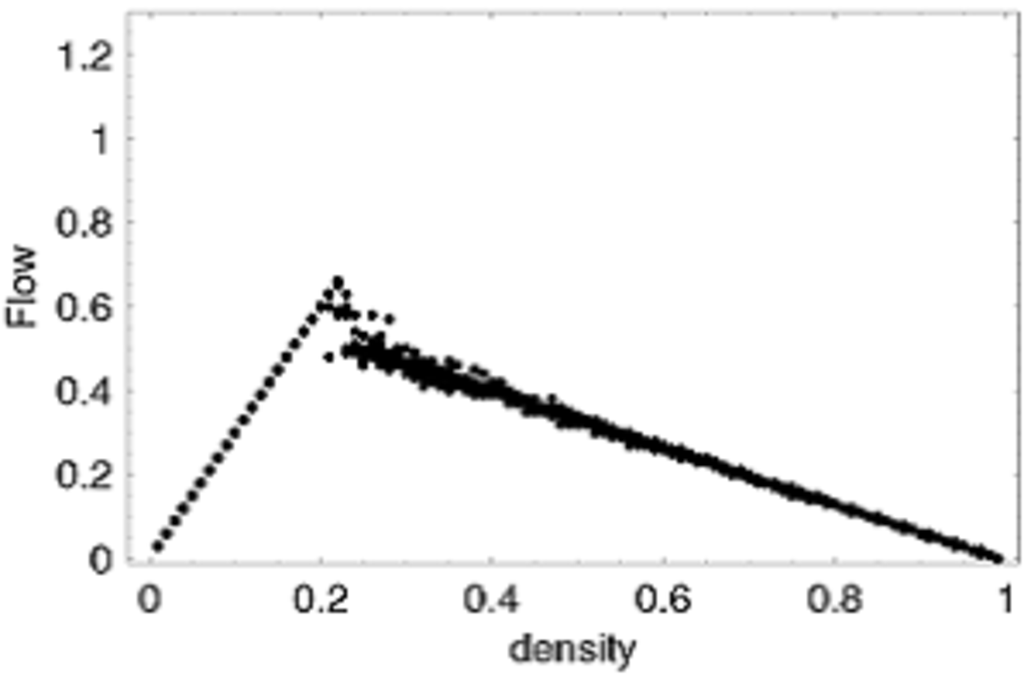}\\
(b)$q = 0.5$, $r = 0.0$\\
  \end{center}
 \end{minipage}
&
 \begin{minipage}{0.30\hsize}
  \begin{center}
   \includegraphics[width=\hsize]{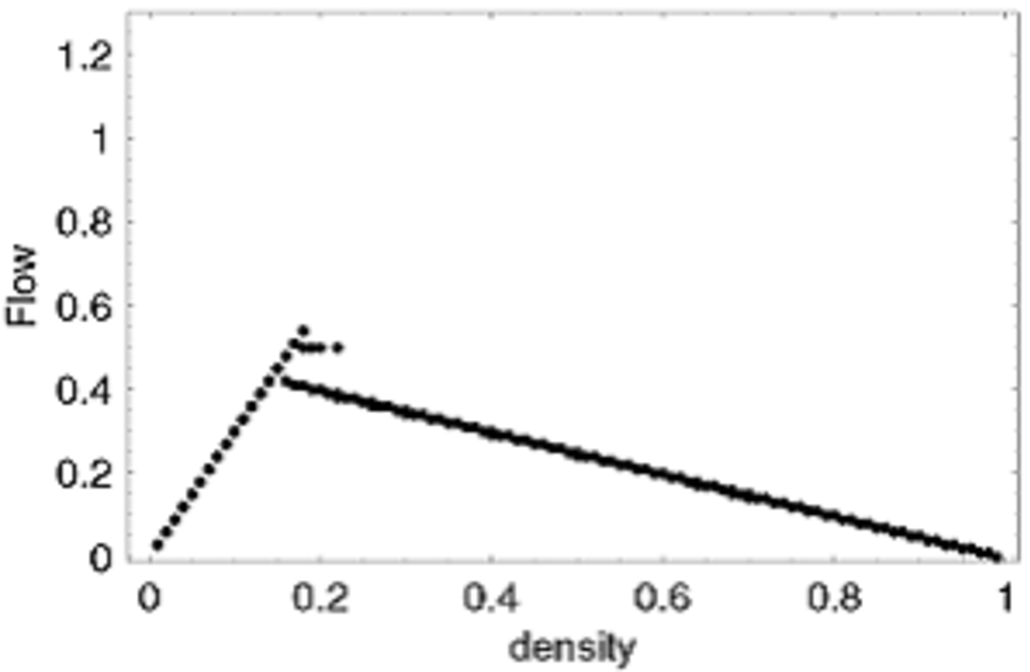}\\
(c)$q = 1.0$, $r = 0.0$\\
  \end{center}
 \end{minipage}\\ \\
 \begin{minipage}{0.30\hsize}
  \begin{center}
   \includegraphics[width=\hsize]{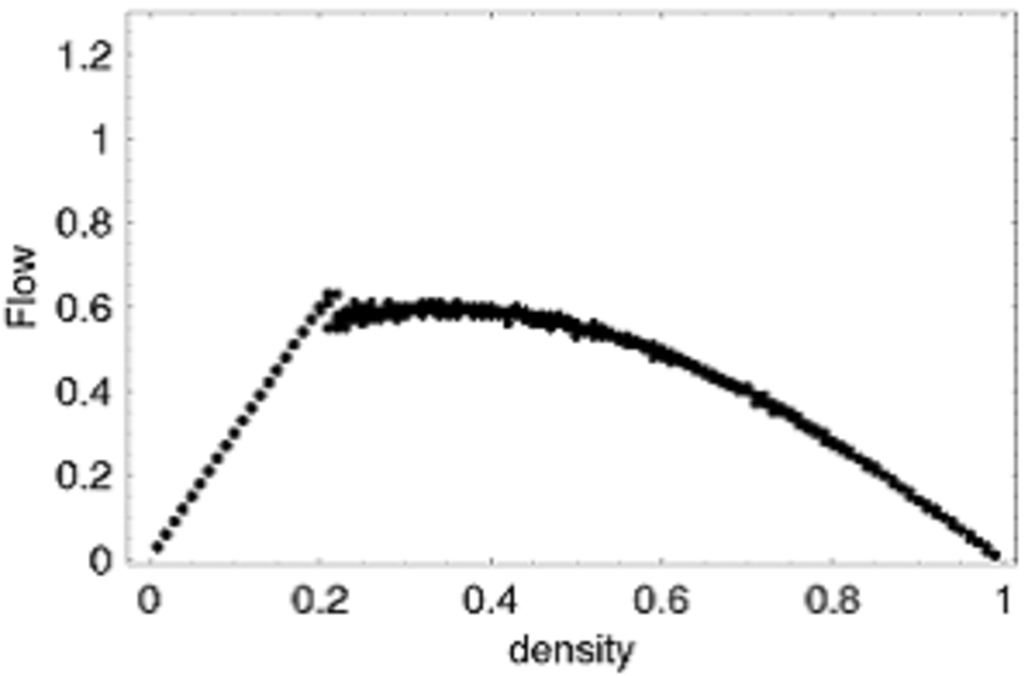}\\
(d)$q = 0.0$, $r = 0.5$\\
  \end{center}
 \end{minipage}
&
 \begin{minipage}{0.30\hsize}
  \begin{center}
   \includegraphics[width=\hsize]{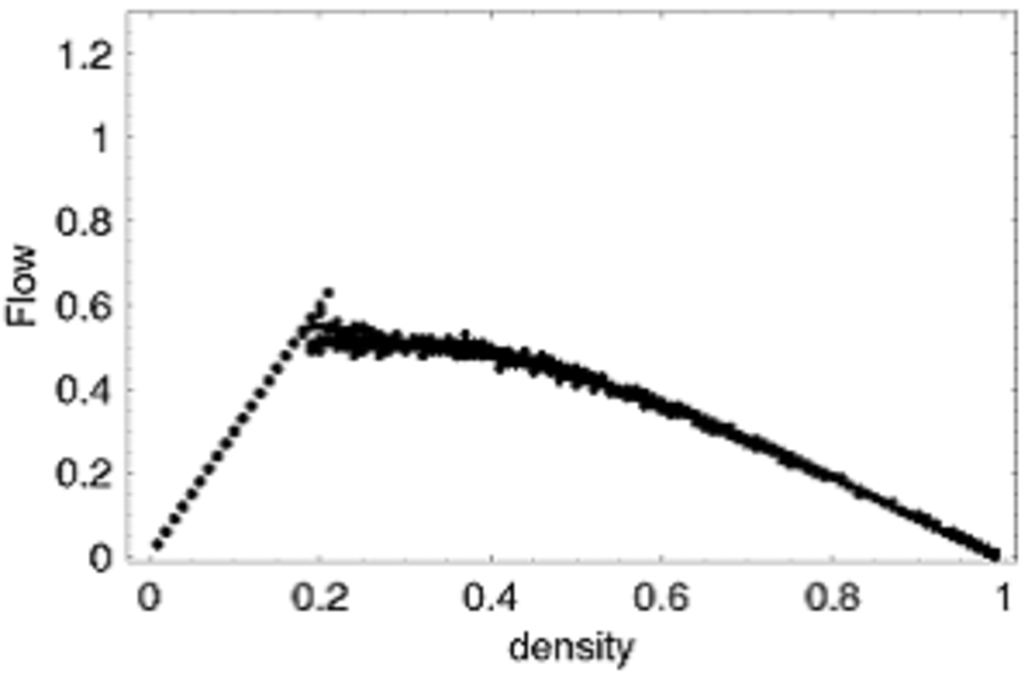}\\
(e)$q = 0.5$, $r = 0.5$\\
  \end{center}
 \end{minipage}
&
 \begin{minipage}{0.30\hsize}
  \begin{center}
   \includegraphics[width=\hsize]{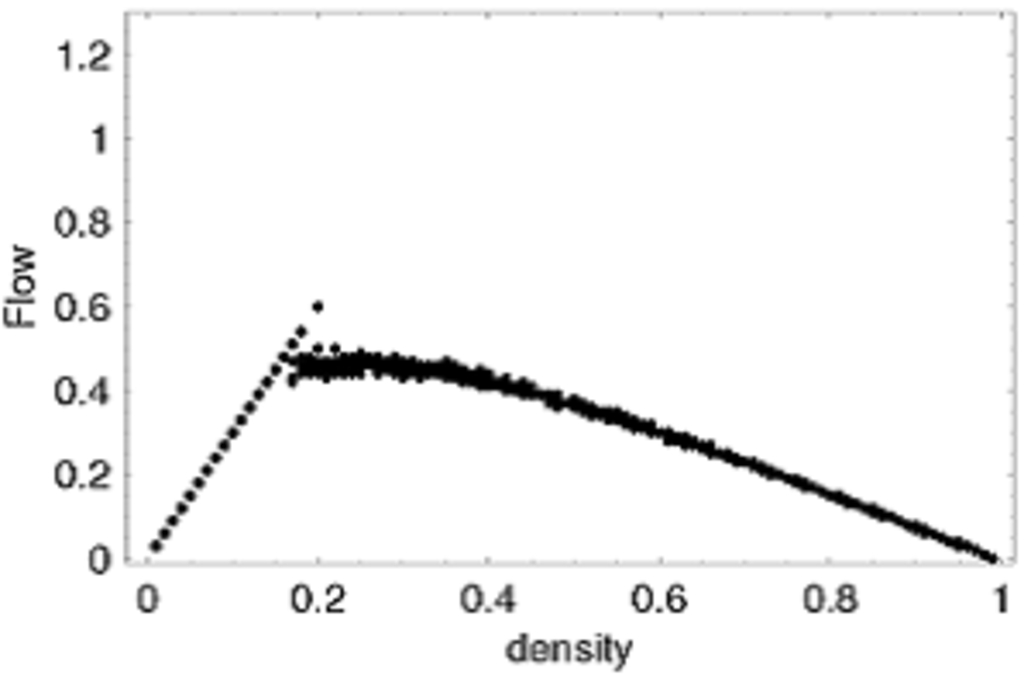}\\
(f)$q = 1.0$, $r = 0.5$\\
  \end{center}
 \end{minipage}\\ \\
 \begin{minipage}{0.30\hsize}
  \begin{center}
   \includegraphics[width=\hsize]{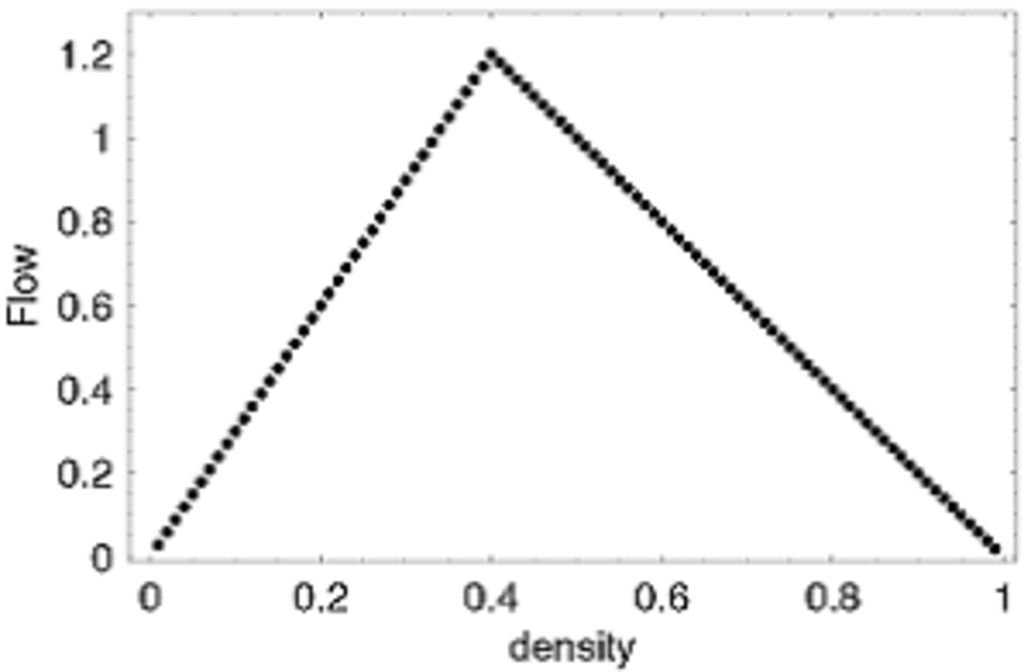}\\
(g)$q = 0.0$, $r = 1.0$\\
  \end{center}
 \end{minipage}
&
 \begin{minipage}{0.30\hsize}
  \begin{center}
   \includegraphics[width=\hsize]{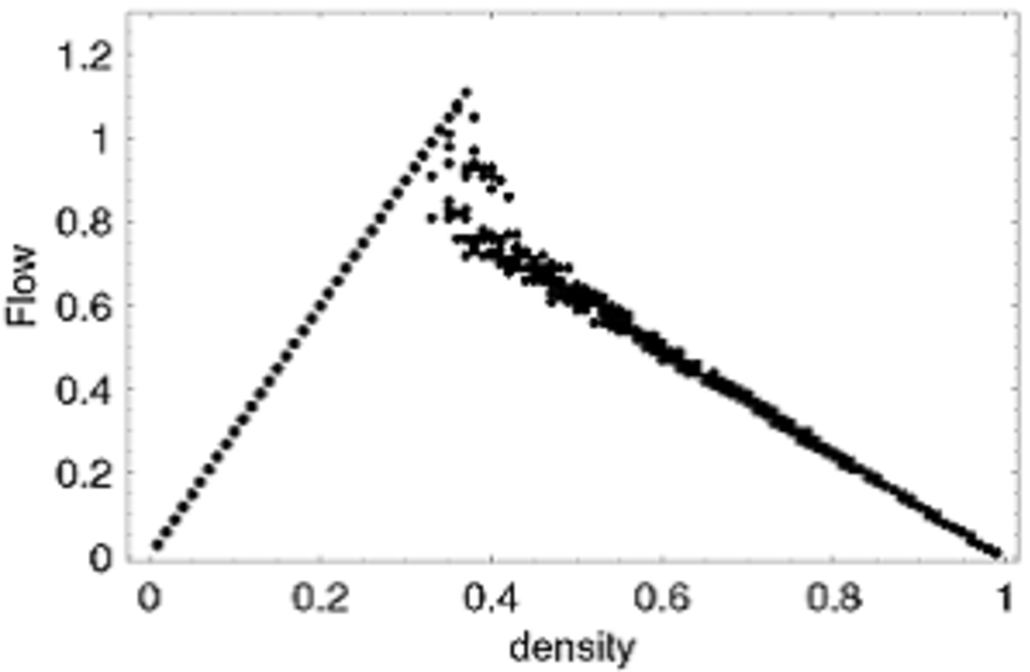}\\
(h)$q = 0.5$, $r = 1.0$\\
  \end{center}
 \end{minipage}
&
 \begin{minipage}{0.30\hsize}
  \begin{center}
   \includegraphics[width=\hsize]{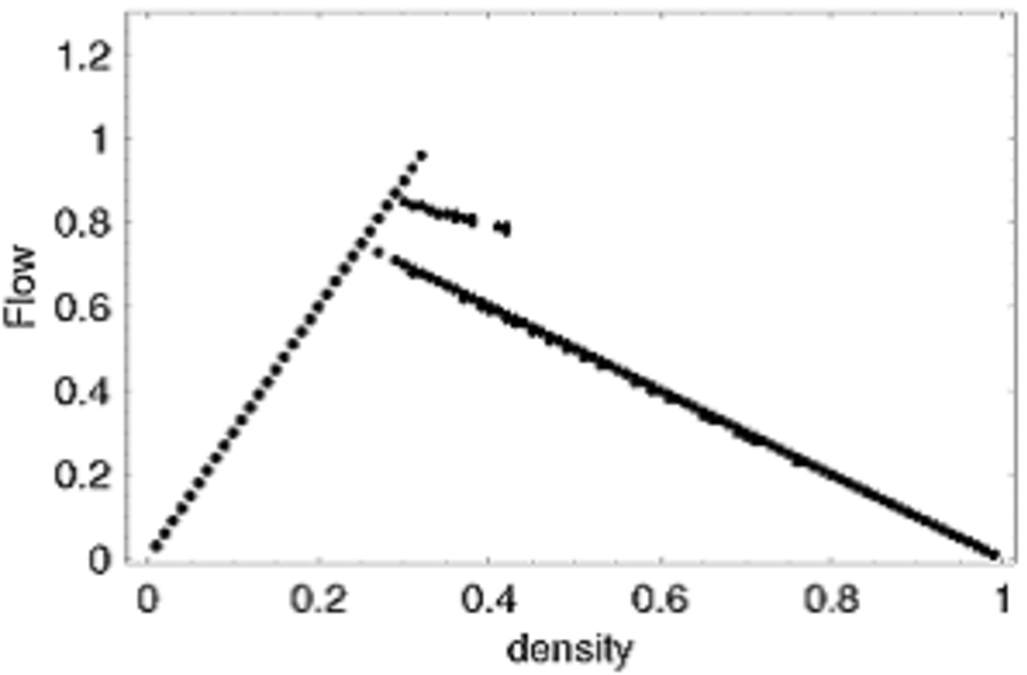}\\
(i)$q = 1.0$, $r = 1.0$\\
  \end{center}
 \end{minipage}
\end{tabular}
\end{center}
\caption{Fundamental diagrams of S-NFS model with $p = 1.0$, 
$L = 100$ and $V_{\max} = 3$.
Contrary to the cases with $V_{\max} = 1$, metastable states 
can remain even for random initial distributions.}
\label{fig:S-NFS_FD3}
\end{figure}

In this section, we consider fundamental diagrams of S-NFS model
on periodic boundary condition (PBC).
Figure \ref{fig:S-NFS_FD} shows
fundamental diagrams with maximum speed $V_{\max} = 1$.
The parameter sets (a), (c), (g), and (i),
correspond to Rule-$184$, SlS model, QS model with $S = 2$,
and NFS model respectively.
Figure \ref{fig:S-NFS_FD3} shows
those with maximam speed $V_{\max} = 3$.
The case (a) is mFI model with $V_{\max} = 3$ and
(i) is NFS model.

The following two points are worthwhile to be mentioned;
\begin{description}
 \item[(1)] Although free flow and jamming lines are both 
straight for $r = 0$ or $r = 1.0$, 
the shape of jamming lines are roundish for $r = 0.5$
(see figures \ref{fig:S-NFS_FD}-(d) or \ref{fig:S-NFS_FD3}-(d),(e),(f))
like ASEP with random braking effect ($p < 1.0$).
The origin of this behavior is as follows:
Two successive cars can move simultaneously as long as $S = 2$.
However, once $S$ of a car moving rear changes to $1$,
this car must stop.
Hence random change of $S$ plays a similar role as random braking.
 \item[(2)] As is seen in figure \ref{fig:S-NFS_FD3}-(h),
metastable states ``stably'' exist even in the presence of stochastic effects,
although in many other models
metastable states soon vanish due to any perturbations.
There are also some new branches around the critical density in our model.
As seen in figure \ref{fig:ActualData},
these new branches may be related to the wide scattering area around
the density $60$--$125$ ($\mbox{vehicles} / \mbox{km}$) in the observed data.
Thus we believe we have successfully reproduce the metastable branches
observed ``stably'' in the empirical data.
\end{description}

\section{Phase diagram}
\begin{figure}[bp]
\begin{center}
\input{figure5.tex}\\[4mm]
\caption{Updating rules for an open boundary condition
of S-NFS model when $V_{\max} = 1$.
(a): Update rules \mbox{(\ref{eq:S-NFS1})}--\mbox{(\ref{eq:S-NFS6})} are
applied to cars within this region.
(b): The rule of slow-to-start \mbox{(\ref{eq:S-NFS2})} is effective only
when $0 \leq x_i^{t-1}$ and $x_{i+S}^{t-1} \leq L-1$.}
\label{fig:S-NFS_OBC}
\end{center}
\end{figure}
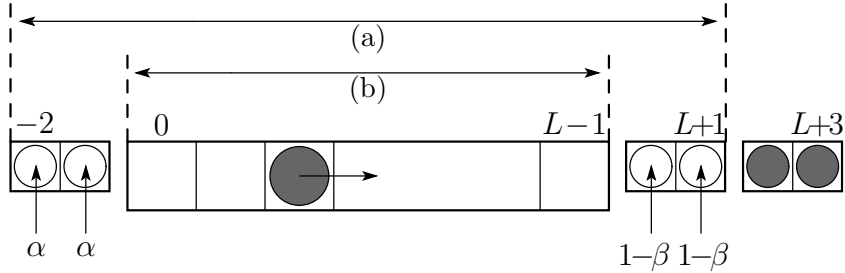

In this section, we consider phase diagrams of S-NFS model
with open boundary condition (OBC) when $V_{\max} = 1$.
Figure \ref{fig:S-NFS_OBC} shows our update rules for OBC:
\begin{description}
 \item[1)] We put two cells at the sites $-2, -1$.
At each site a car is injected with probability $\alpha$.
The car's velocity is set to $1$.
 \item[2)] We put cells at the sites $L, L+1$, and at each site
a car is injected with probability $1-\beta$
(an outflow of a car from right end is obstructed by these cars).
It's velocity is set to $0$.
 \item[3)] Besides, we put cells at the sites $L+2, L+3$.
At these two sites cars are always injected
(this rule means that for cars at sites $-2, \ldots, L+1$, there
exist at least two preceding cars at any time).
It's velocity is set to $0$.
 \item[4)] We apply updating rule of S-NFS to cars
at sites $-2, \ldots, L+1$.
However slow-to-start effect \mbox{(\ref{eq:S-NFS2})} apply
only if $0 \leq x_i^{t-1}$ and $x_{i+S}^{t-1} \leq L-1$.
 \item[5)] We remove all cars at $-2, -1$ and $L, \ldots, L+3$
at the end of each time step.
\end{description}
These rules are devised in order to avoid unnatural traffic
jam caused by boundary conditions.
Here, the rule {\bf 3)} is needed in order that
for cars at sites $-2, \ldots, L+1$ there always exist
at least two preceding cars.
We need to apply S-NFS update rules to cars at sites $L, L+1$
because otherwise the velocity of the car at site $L-1$ is
determined irrepective of the state at $L+1$ even when $S = 2$.

\begin{figure}[ptbh]
\begin{center}
\begin{tabular}{ccc}
 \begin{minipage}{0.30\hsize}
  \begin{center}
   \includegraphics[width=\hsize]{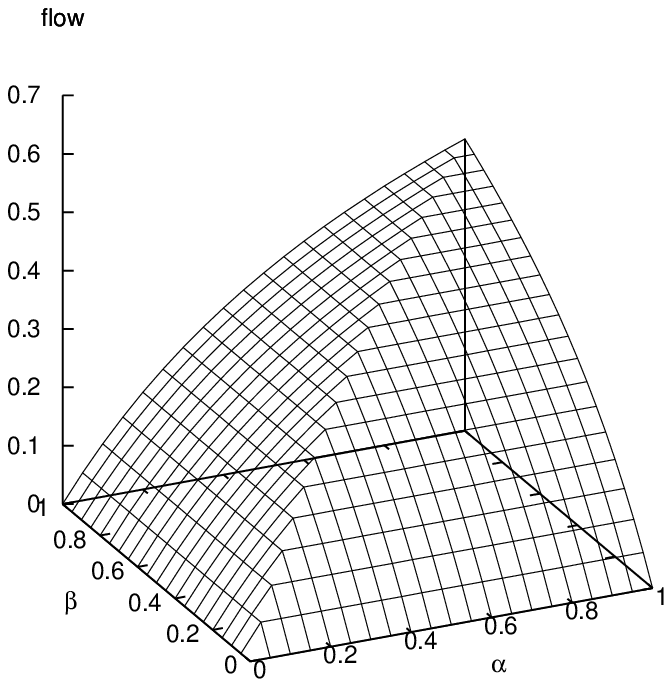}\\
(a)flow-$\alpha$-$\beta$ diagram\\
  \end{center}
 \end{minipage}
&
 \begin{minipage}{0.30\hsize}
  \begin{center}
   \includegraphics[width=\hsize]{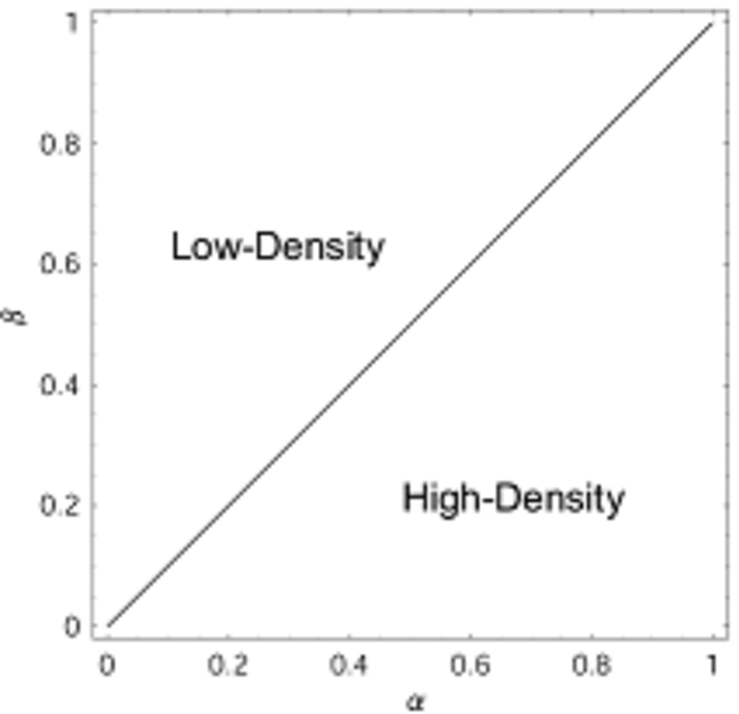}\\
(b)$\alpha$-$\beta$ phase diagram\\
  \end{center}
 \end{minipage}
\end{tabular}
\end{center}
\caption{Definition of flow-$\alpha$-$\beta$ diagram of ASEP
($p = 1.0, q = r = 0.0$)
and a corresponding $\alpha$-$\beta$ phase diagram.
The line of the first-order phase transition is given by $\alpha = \beta$.}
\label{fig:ASEP_PD}
\end{figure}

\begin{figure}[btp]
\begin{center}
\begin{tabular}{ccc}
 \begin{minipage}{0.30\hsize}
  \begin{center}
   \includegraphics[width=\hsize]{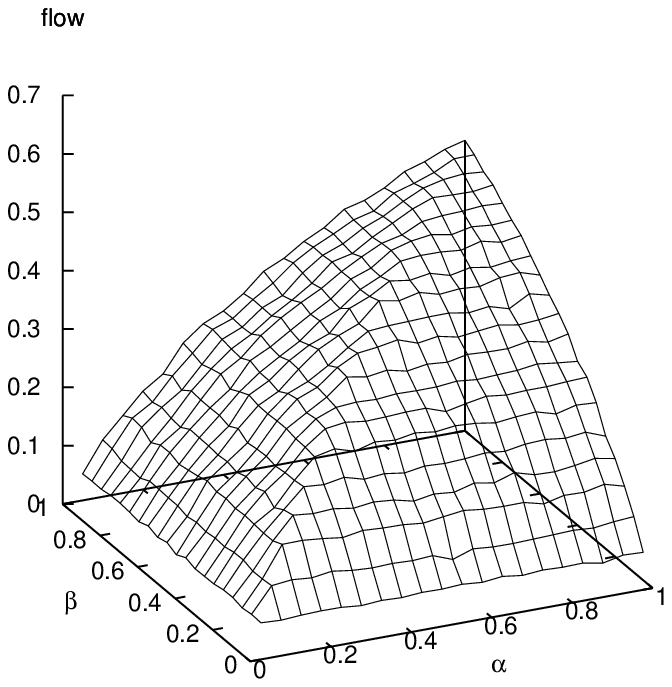}\\
(a)$q = 0.0$, $r = 0.0$\\
  \end{center}
 \end{minipage}
&
 \begin{minipage}{0.30\hsize}
  \begin{center}
   \includegraphics[width=\hsize]{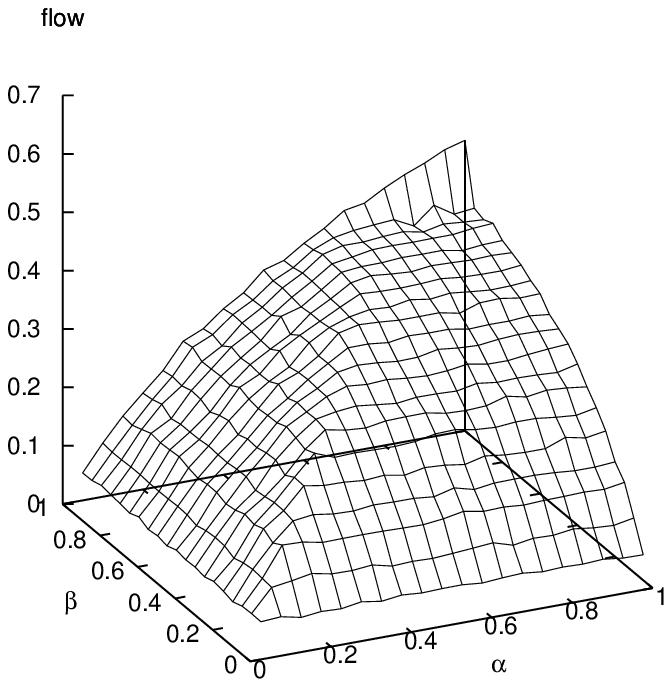}\\
(b)$q = 0.5$, $r = 0.0$\\
  \end{center}
 \end{minipage}
&
 \begin{minipage}{0.30\hsize}
  \begin{center}
   \includegraphics[width=\hsize]{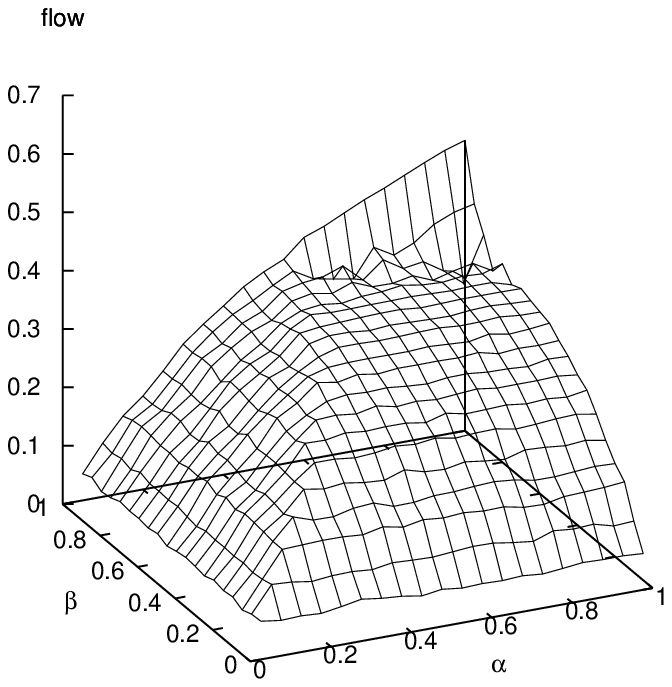}\\
(c)$q = 1.0$, $r = 0.0$\\
  \end{center}
 \end{minipage}\\ \\
 \begin{minipage}{0.30\hsize}
  \begin{center}
   \includegraphics[width=\hsize]{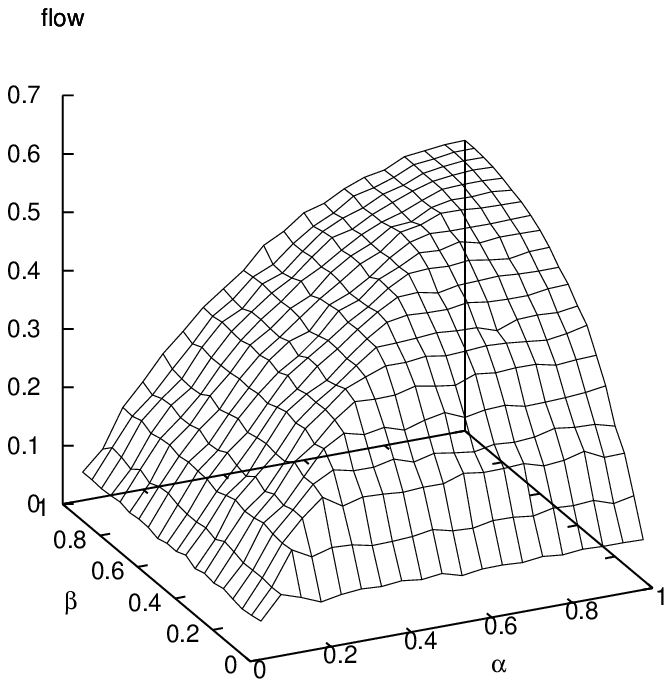}\\
(d)$q = 0.0$, $r = 0.5$\\
  \end{center}
 \end{minipage}
&
 \begin{minipage}{0.30\hsize}
  \begin{center}
   \includegraphics[width=\hsize]{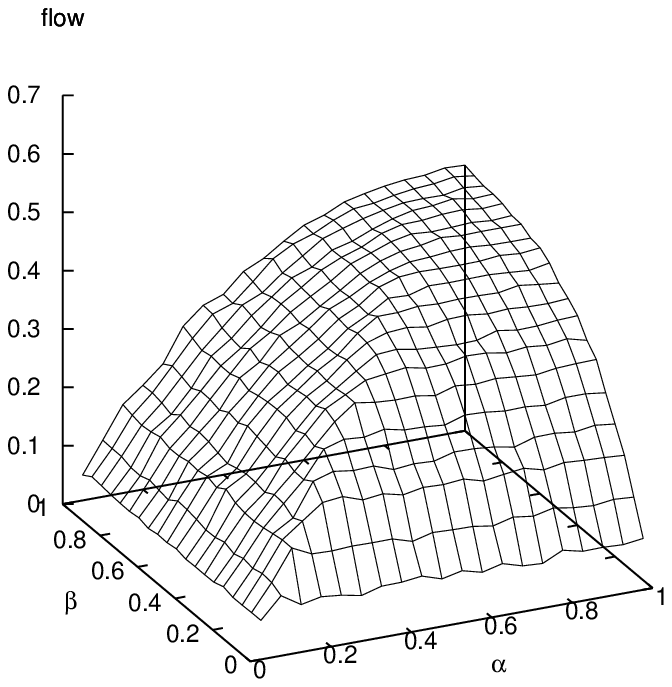}\\
(e)$q = 0.5$, $r = 0.5$\\
  \end{center}
 \end{minipage}
&
 \begin{minipage}{0.30\hsize}
  \begin{center}
   \includegraphics[width=\hsize]{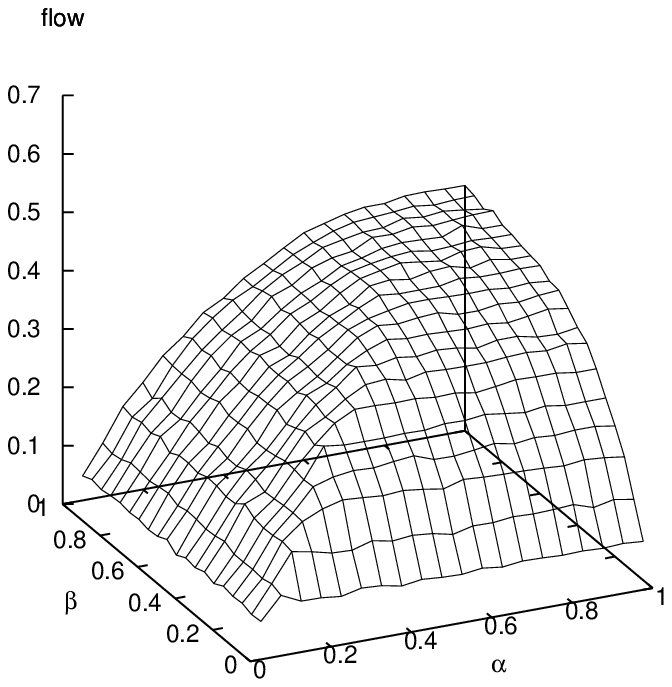}\\
(f)$q = 1.0$, $r = 0.5$\\
  \end{center}
 \end{minipage}\\ \\
 \begin{minipage}{0.30\hsize}
  \begin{center}
   \includegraphics[width=\hsize]{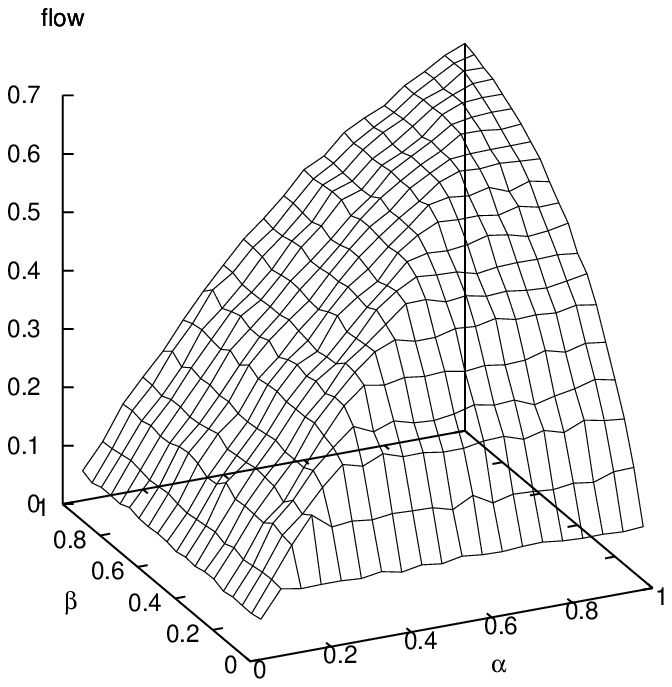}\\
(g)$q = 0.0$, $r = 1.0$\\
  \end{center}
 \end{minipage}
&
 \begin{minipage}{0.30\hsize}
  \begin{center}
   \includegraphics[width=\hsize]{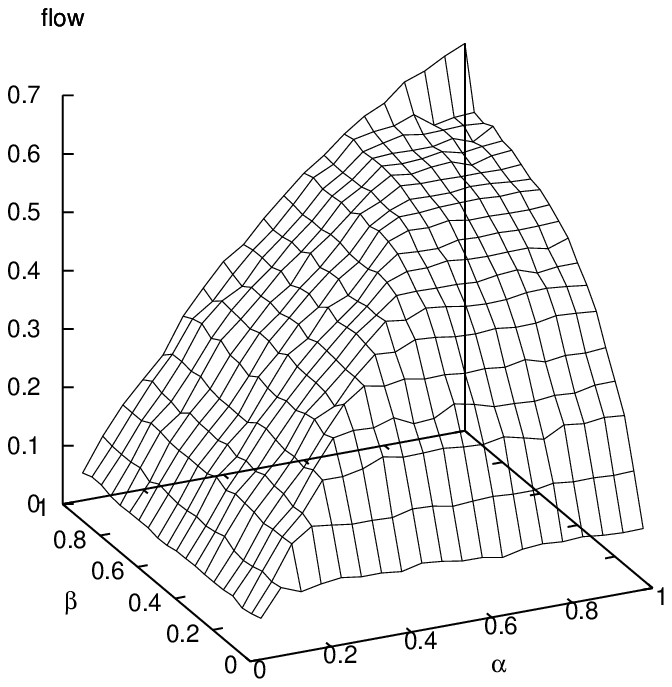}\\
(h)$q = 0.5$, $r = 1.0$\\
  \end{center}
 \end{minipage}
&
 \begin{minipage}{0.30\hsize}
  \begin{center}
   \includegraphics[width=\hsize]{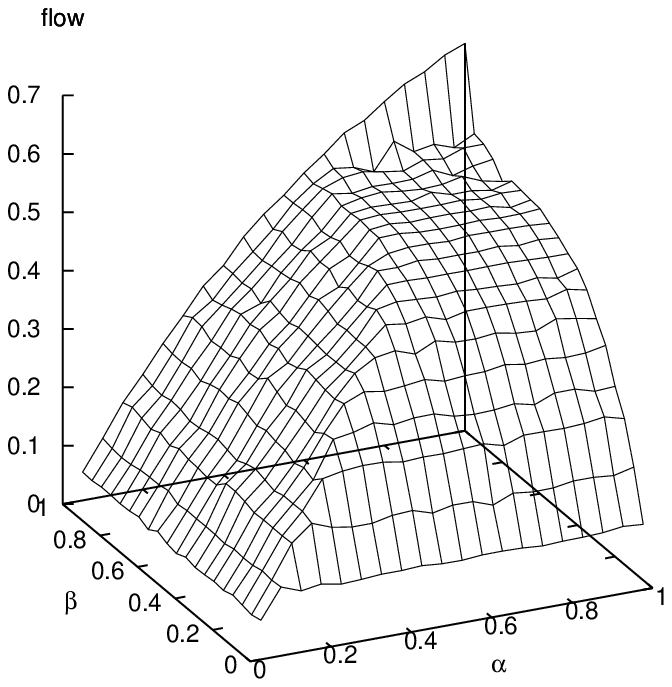}\\
(i)$q = 1.0$, $r = 1.0$\\
  \end{center}
 \end{minipage}
\end{tabular}
\end{center}
\caption{Flow-$\alpha$-$\beta$ diagrams of S-NFS model with $p = 1.0$
and $V_{\max} = 1$.
Note that flow suddenly rises near $\beta = 1.0$ in (b), (c), (h), and (i)
because cars never take the slow-to-start mode even 
when $q \neq 0$.}
\label{fig:S-NFS_fab}
\end{figure}

With use of this boundary condition, we can calculate
flow-$\alpha$-$\beta$ diagrams (see figure \ref{fig:S-NFS_fab}) form
which phase diagrams are derived (see figures \ref{fig:S-NFS_PD}).
The phase diagram of ASEP has been already known
(see figure \ref{fig:ASEP_PD}).
In this diagram, the whole $\alpha$-$\beta$ region is divided into two phases
in the case $p = 1.0$:
low-density (LD) phase which is ``$\alpha$ controlled phase'';
and high-density (HD) phase which is ``$\beta$ controlled phase''.
On the phase transition curve, the effect of $\alpha$ is balanced with
that of $\beta$,
and then these two phases coexist.
The flow-$\alpha$-$\beta$ diagrams of S-NFS model are considerably
different form the result of ASEP (see figure \ref{fig:S-NFS_fab}).
The difference will be discussed in sec.\ref{sec:PTL}.
We restrict ourselves to the case $V_{\max} = 1$ and $p = 1.0$
in the following.

\section{Phase transition curve}
\label{sec:PTL}
In this section, let us derive an analytic expression of
the first-order phase transition curve of S-NFS model
by combining approximate flow-density relations
at boundaries and fundamental diagrams.
We extend the method proposed in articles \cite{DLG_ASEP_PD, Cecile}.
The method consists of three steps: 
first, we relate the flow and density near the boundaries of the system in OBC;
next, in fundamental diagrams of PBC, we calculate the gradient of
free-flow line and jamming line;
finally we can obtain phase diagrams for OBC from the above results.

\subsection{Relation between flow and density near the boundaries}
\begin{figure}[ht]
\begin{center}
\begin{tabular}{llll}
 \begin{minipage}{0.2\hsize}
 \begin{center}
  \input{figure8a.tex}\\
(a)
  \end{center}
 \end{minipage}
 &
 \begin{minipage}{0.2\hsize}
  \begin{center}
  \input{figure8b.tex}\\
(b)
  \end{center}
 \end{minipage}
 &
 \begin{minipage}{0.2\hsize}
  \begin{center}
  \input{figure8c.tex}\\
(c)
  \end{center}
 \end{minipage}
 &
 \begin{minipage}{0.2\hsize}
  \begin{center}
  \input{figure8d.tex}\\
(d)
  \end{center}
 \end{minipage}
\end{tabular}
\end{center}
\caption{Configurations of cars contributing to flow at left or 
right ends. Other configurations do not generate flow at both ends.}
\label{fig:boundaryFlow}
\end{figure}
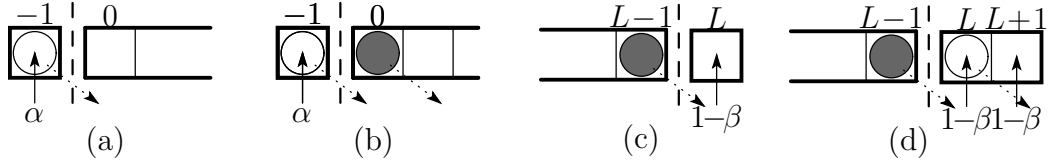

First, we calculate each flow $J_l$ and $J_r$ at left or right end of
this system.
Configurations in figure \ref{fig:boundaryFlow} contribute flow
at each boundaries.
Then with use of mean-field approach,
the probability of the configration (a) is $c_{(-1)}[1-c_{(0)}]$.
Similarly the probabilities of (b), (c), and (d) are
$rc_{(-1)}c_{(0)}[1-c_{(1)}]$, $c_{(L-1)}[1-c_{(L)}]$,
and $rc_{(L-1)}c_{(L)}[1-c_{(L+1)}]$ respectively.
Here, $c_{(j)}$ denotes the probability of finding a car at the site $j$.
Because $c_{(-1)} = \alpha$ and $c_{(L)} = c_{(L+1)} = 1 - \beta$,
we get 
\begin{equation}
\label{eq:ends_flow}
\left\{
\begin{array}{l}
J_l = \alpha (1-c_{(0)})(1+rc_{(0)})\\
J_r = \beta [1+r(1-\beta)]c_{(L-1)}
\end{array}
\right.
\end{equation}
where we assume $c_{(0)} = c_{(1)}$.
Note, we ignored possibility that a car at the site $0$ can not move
by slow-to-start effect in (b).

\subsection{Approximate expression of fundamental diagrams}
Next we approximate free-flow line and jamming line of PBC
by straight lines as
\begin{equation}
\label{eq:flow-rho}
\left\{
\begin{array}{ll}
J_f = \rho = c_{(0)} & \mbox{(free-flow line)}\\
J_j = x(1-\rho) = x(1-c_{(L-1)}) \quad & \mbox{(jamming line)}
\end{array}
\right.
,
\end{equation}
where $J_f$ and $J_j$ denote the flow of free-flow and jamming phases,
and $x$ is a magnitude of the gradient of jamming line.
Here, we need to get the relations between $x$ and $q$ or $r$.
The parameter $x$ is related to the velocity of
backward moving traffic clusters which we will calculate with use of
mean-field approach.

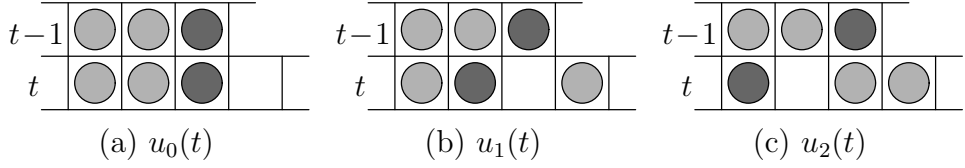
\begin{figure}[ht]
\begin{center}
\begin{tabular}{lll}
 \begin{minipage}{0.25\hsize}
 \begin{center}
  \input{figure9a.tex}\\
(a) $u_0(t)$
  \end{center}
 \end{minipage}
 &
 \begin{minipage}{0.25\hsize}
  \begin{center}
  \input{figure9b.tex}\\
(b) $u_1(t)$
  \end{center}
 \end{minipage}
 &
 \begin{minipage}{0.25\hsize}
  \begin{center}
  \input{figure9c.tex}\\
(c) $u_2(t)$
  \end{center}
 \end{minipage}
\end{tabular}
\end{center}
\caption{The connection between parameter $u_g(t)$ and configurations of cars.}
\label{fig:backward_moving}
\end{figure}

\begin{figure}[ht]
\begin{center}
\input{figure10.tex}\\
\caption{The case that $u_2(t)$ becomes $u_1(t+1)$.}
\label{fig:backward_example}
\end{center}
\end{figure}
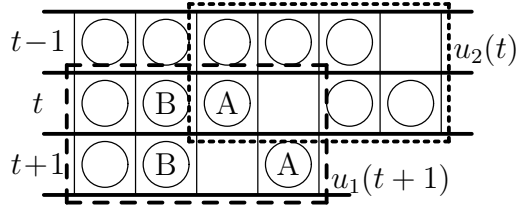

\begin{figure}[hbpt]
\begin{center}
\begin{tabular}{ccc}
 \begin{minipage}{0.45\hsize}
  \begin{center}
   \includegraphics[width=\hsize]{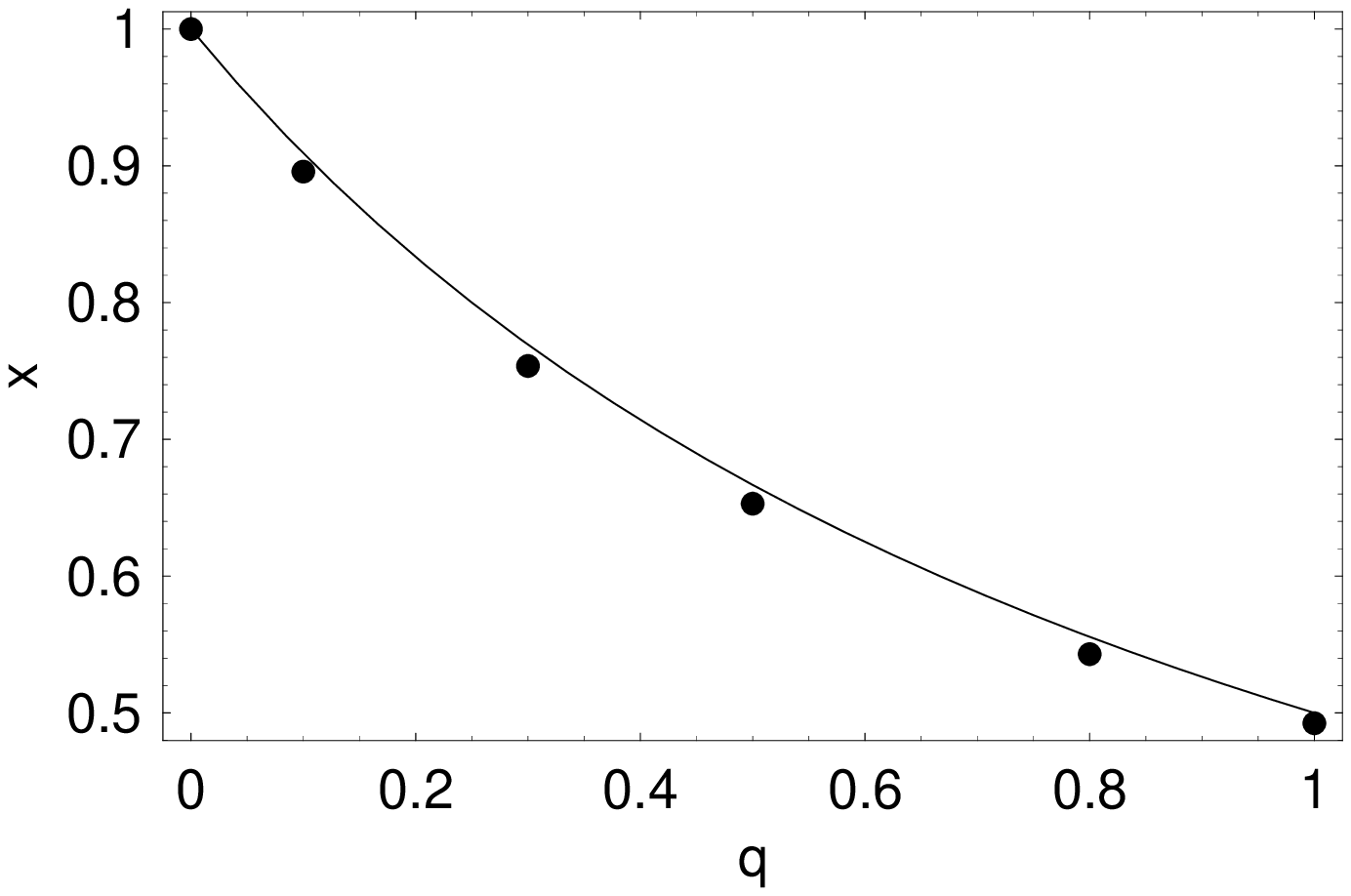}\\
(a)$x$-$q$ diagram ($r = 0.0$)\\
  \end{center}
 \end{minipage}
&
 \begin{minipage}{0.45\hsize}
  \begin{center}
   \includegraphics[width=\hsize]{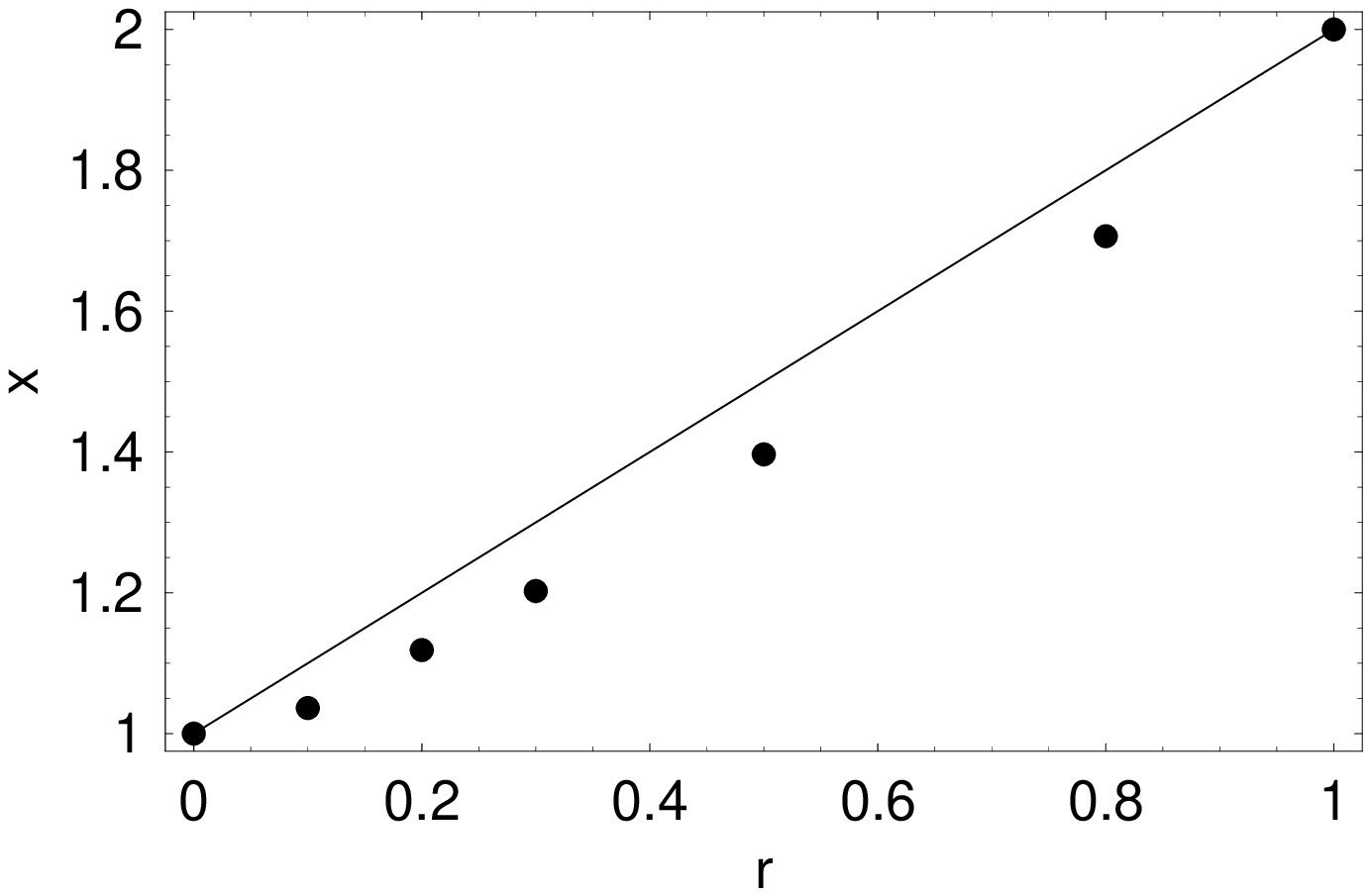}\\
(b)$x$-$r$ diagram ($q = 0.0$)\\
  \end{center}
 \end{minipage}
\end{tabular}
\end{center}
\caption{The relation between the gradient of jamming line $x$ 
and probabilities $q$ or $r$.
The solid line denotes the approximate analytic result,
and points denote results of numerical simulations.
There seem some discrepancies in (b).
The reason is that we assume the shape of jamming line
in the fundamental diagram to be straight although it is actually roundish.}
\label{fig:x-qr}
\end{figure}

Figure \ref{fig:backward_moving} shows the possible configurations
at the right end of clusters.
The symbols $u_g(t)\;$ $g \in \{0,\; 1,\; 2\}$ mean the probability of
each configurations at time $t$. The subscript $g$ denotes
the gradient of each configurations.
For example, let us consider the probability that the configuration
figure \ref{fig:backward_moving} (c) becomes
(b) in the next time step, $u_2(t) \rightarrow u_1(t+1)$
(see figure \ref{fig:backward_example}).
In order that (b) occurs, the slow-to-start effect is not active
for the car A (the probability $1-q$).
The car B does not move when (i) the driver's perspective is not active
(the probability $1-r$) or (ii) although the driver's perspective is active,
the car B still halts due to slow-to-start effect (the probability $q r$).
Thus we get the term $u_2(t)(1 - q)(1 - r + q r)$ which expresses
the probability that $u_2(t)$ becomes $u_1(t+1)$.
Considering other possibilities in the same way,
we get the following recursion relations
\begin{eqnarray}
\label{eq:MFA_u_eq1}
\fl
u_0(t+1) &=& u_1(t)q(1-r)+u_2(t)q\\
\label{eq:MFA_u_eq2}
\fl
u_1(t+1) &=& u_0(t)(1-r) + u_1(t)(1-r+qr)(1-q+qr)+u_2(t)(1-q)(1-r+qr)\\
\label{eq:MFA_u_eq3}
\fl
u_2(t+1) &=& u_0(t)r + u_1(t)r(1-q)(1-q+qr)+u_2(t)r(1-q)^2.
\end{eqnarray}
We solve the equations \mbox{(\ref{eq:MFA_u_eq1})},
\mbox{(\ref{eq:MFA_u_eq2})}, and \mbox{(\ref{eq:MFA_u_eq3})}
for the steady state $u_g(t+1) = u_g(t) = u_g$, and we get
\begin{eqnarray}
u_0 &=&
 \frac{q[1-r+(1-q)r^2]}{1-2q^2r^2+q(1-r+r^2)}\\
u_1 &=&
 \frac{1-(1-q+q^2)r}{1-2q^2r^2+q(1-r+r^2)}\\
u_2 &=&
 \frac{[1-q+q^2(1-r)]r}{1-2q^2r^2+q(1-r+r^2)}.
\end{eqnarray}
Thus, we take the expectation $x = 0u_0 + 1u_1 + 2u_2$, and
obtain the answer:
\begin{equation}
\label{eq:answer_x}
x = x(q, r) = \frac{1+r-qr+q^2r-2q^2r^2}{1+q-qr+qr^2-2q^2r^2}.
\end{equation}
The relations between $x$ and $q$ or $r$ obtained from numerical simulations
are plotted and compared with eq.(\ref{eq:answer_x}) in figure \ref{fig:x-qr}.
The theoretical curve (\ref{eq:answer_x}) gives good estimation
of numerical results.

\subsection{Derivation of phase transition curve}
Finally, we derive phase transition curve for OBC from above results.
From \mbox{(\ref{eq:ends_flow})} and \mbox{(\ref{eq:flow-rho})},
we obtain the following equations
\begin{eqnarray}
J_f = J_l &\quad \Longrightarrow \quad& c_{(0)} = \alpha[1-c_{(0)}][1+rc_{(0)}]\\
J_j = J_r &\quad \Longrightarrow \quad& x[1-c_{(L-1)}] = \beta[1+r(1-\beta)]c_{(L-1)},
\end{eqnarray}
which lead to
\begin{eqnarray}
\label{eq:rho_lr}
\displaystyle
c_{(0)} = \frac{\alpha(r-1)-1
 + \sqrt{\alpha^2(1+2r+r^2)+2\alpha(1-r)+1}}{2r\alpha}\\
\label{eq:rho_lr2}
\displaystyle
c_{(L-1)} = \frac{x}{\beta[1+r(1-\beta)]+x}
.
\end{eqnarray}
Using \mbox{(\ref{eq:flow-rho})} and the equation $J_f = J_j$
which is the characteristic of two phase coexistence,
we can get $c_{(0)} = x[1-c_{(L-1)}]$.
Substituting \mbox{(\ref{eq:rho_lr2})} into this equation,
we can express $\beta$ by $q$, $r$, $x$, and $\alpha$:
\begin{eqnarray}
\label{eq:beta-alpha1}
\beta = \frac{1+r}{2r} +
 \frac{\sqrt{[c_{(0)}-x]^2(1+r)^2
  +4rx[c_{(0)}-x]c_{(0)}}}{2r[c_{(0)}-x]}
\end{eqnarray}
where $c_{(0)}$ is written by $\alpha$ and $r$
as \mbox{equation (\ref{eq:rho_lr})}.
This is the approximated curve of the phase boundary.

\begin{figure}[bpt]
\begin{center}
\begin{tabular}{ccc}
 \begin{minipage}{0.30\hsize}
  \begin{center}
   \includegraphics[width=\hsize]{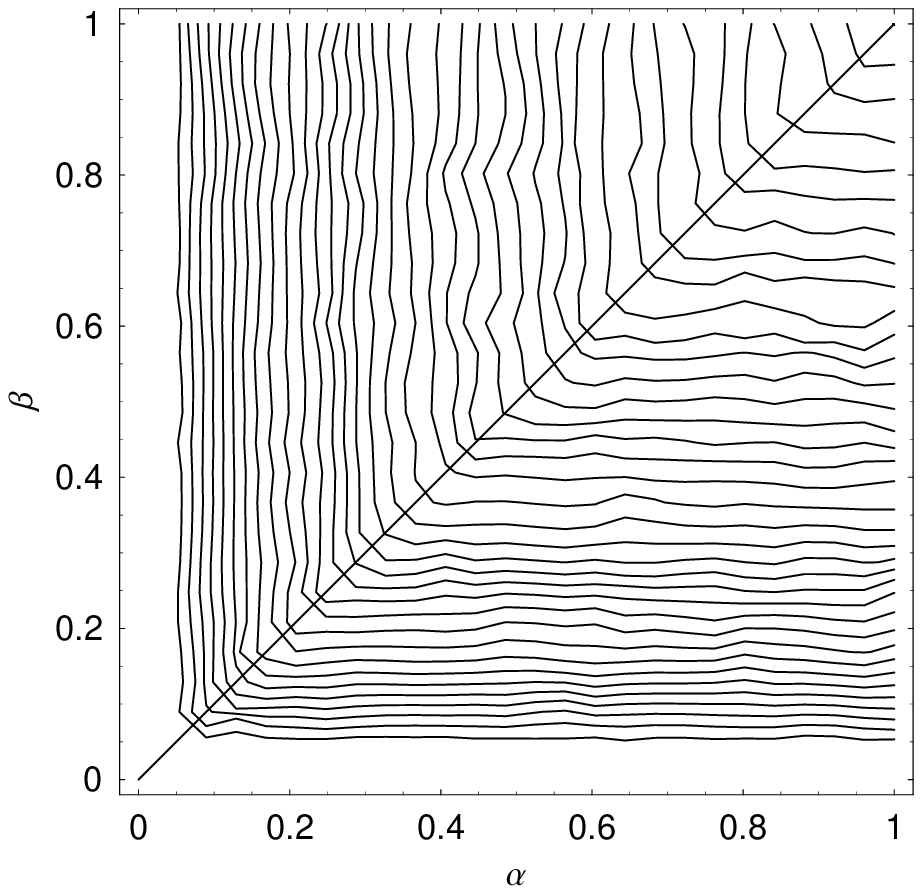}\\
(a)$q = 0.0$, $r = 0.0$\\
  \end{center}
 \end{minipage}
&
 \begin{minipage}{0.30\hsize}
  \begin{center}
   \includegraphics[width=\hsize]{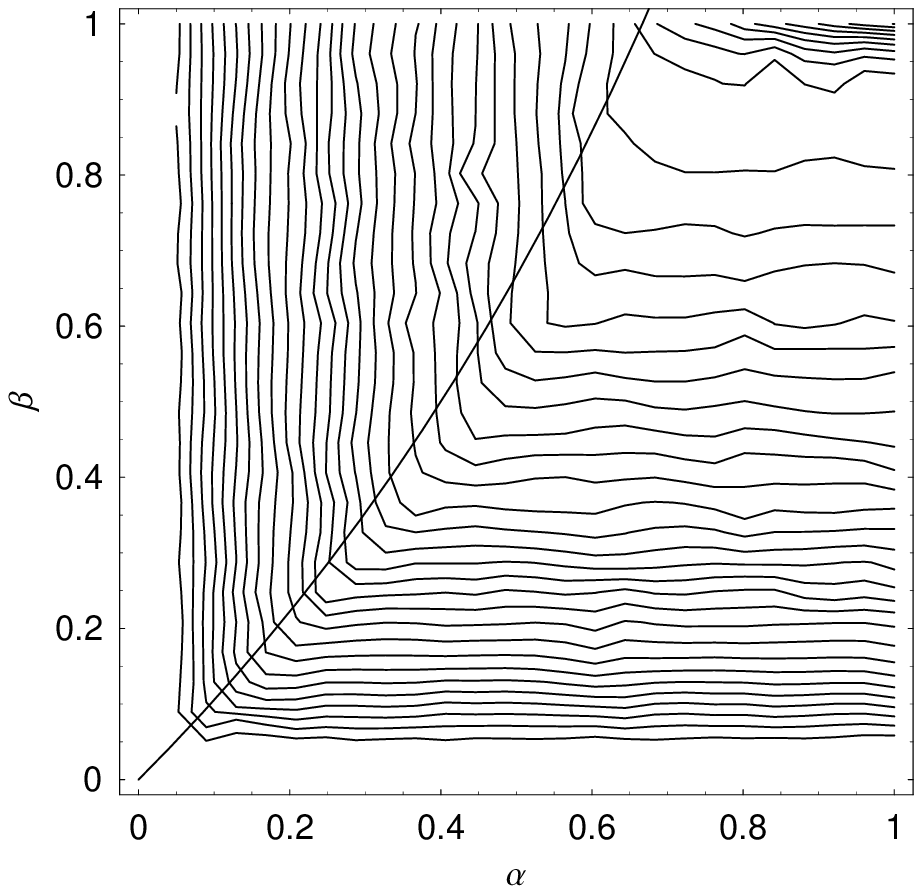}\\
(b)$q = 0.5$, $r = 0.0$\\
  \end{center}
 \end{minipage}
&
 \begin{minipage}{0.30\hsize}
  \begin{center}
   \includegraphics[width=\hsize]{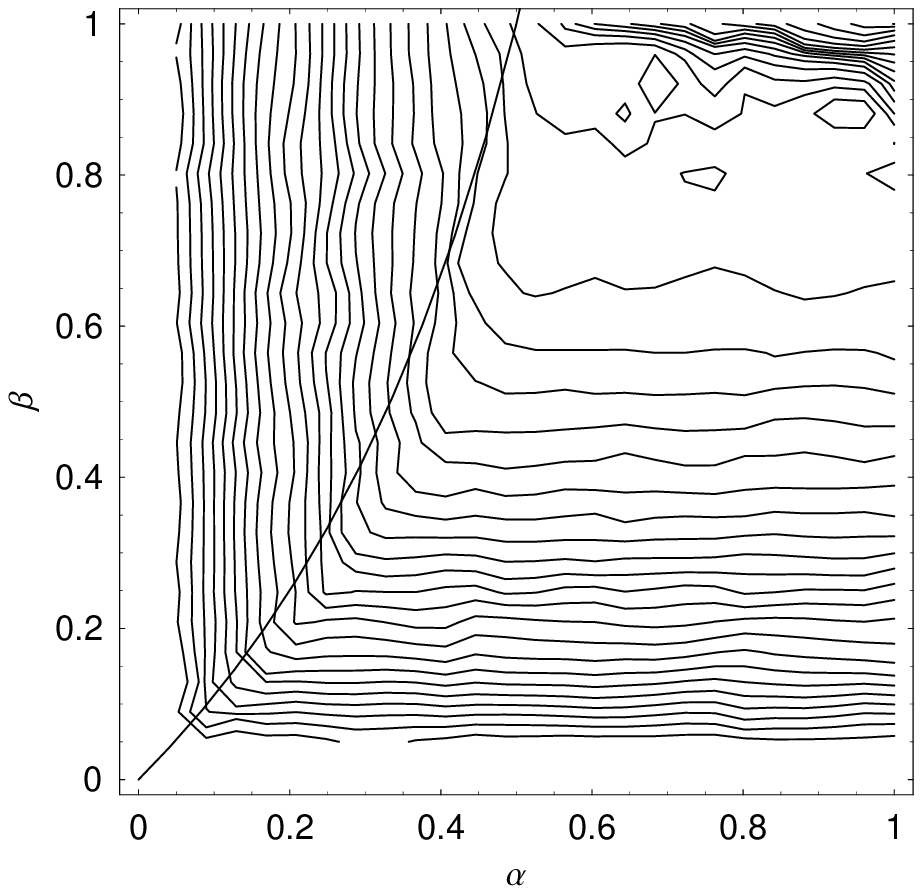}\\
(c)$q = 1.0$, $r = 0.0$\\
  \end{center}
 \end{minipage}\\ \\
 \begin{minipage}{0.30\hsize}
  \begin{center}
   \includegraphics[width=\hsize]{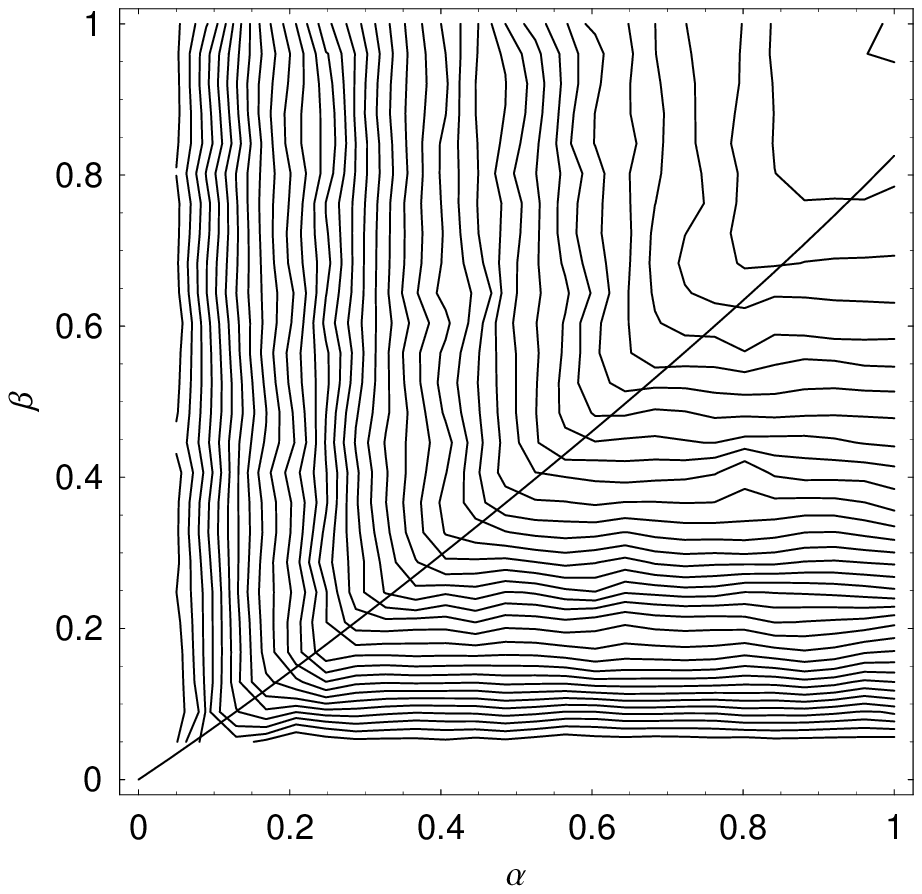}\\
(d)$q = 0.0$, $r = 0.5$\\
  \end{center}
 \end{minipage}
&
 \begin{minipage}{0.30\hsize}
  \begin{center}
   \includegraphics[width=\hsize]{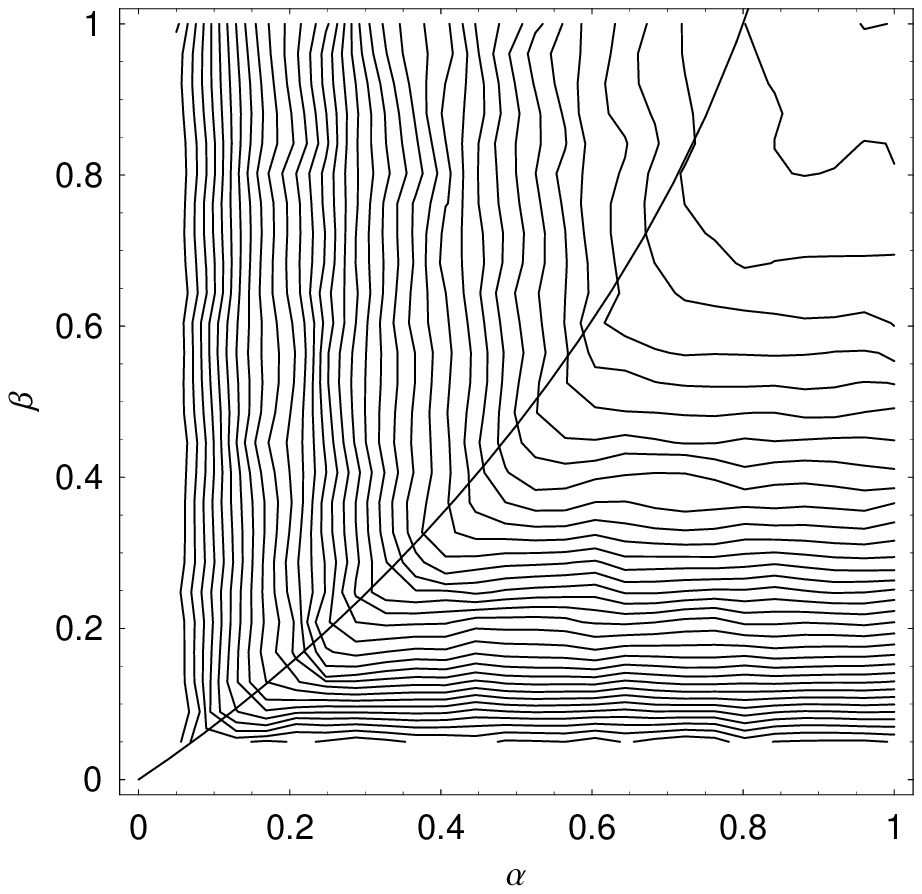}\\
(e)$q = 0.5$, $r = 0.5$\\
  \end{center}
 \end{minipage}
&
 \begin{minipage}{0.30\hsize}
  \begin{center}
   \includegraphics[width=\hsize]{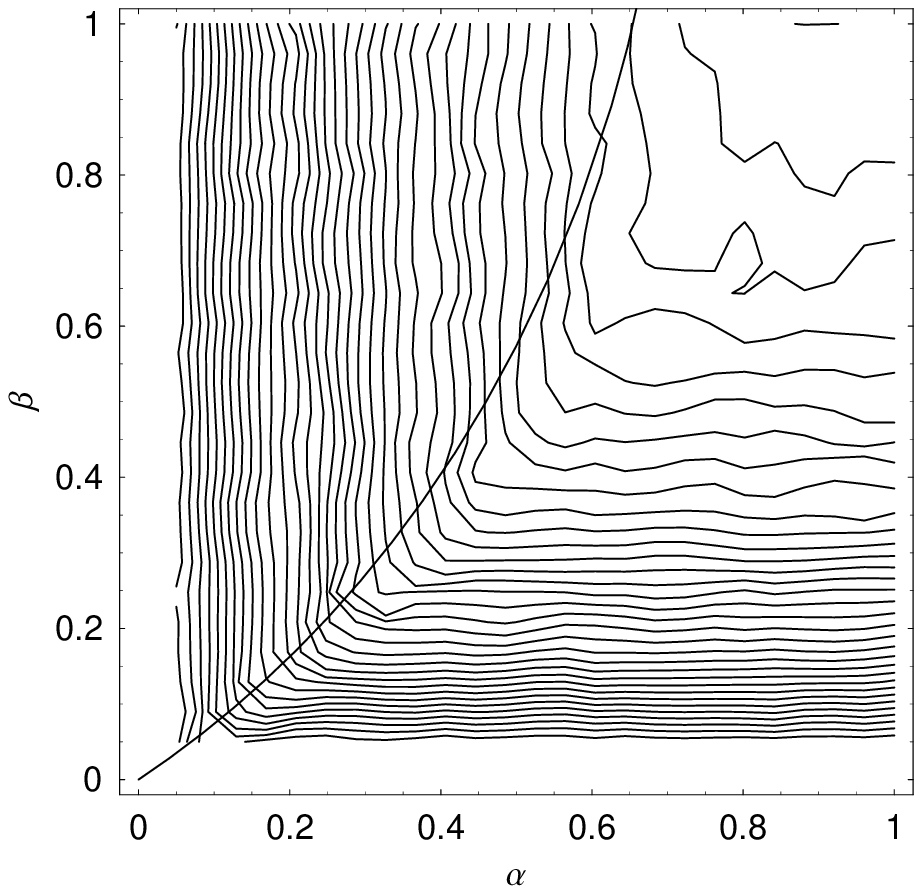}\\
(f)$q = 1.0$, $r = 0.5$\\
  \end{center}
 \end{minipage}\\ \\
 \begin{minipage}{0.30\hsize}
  \begin{center}
   \includegraphics[width=\hsize]{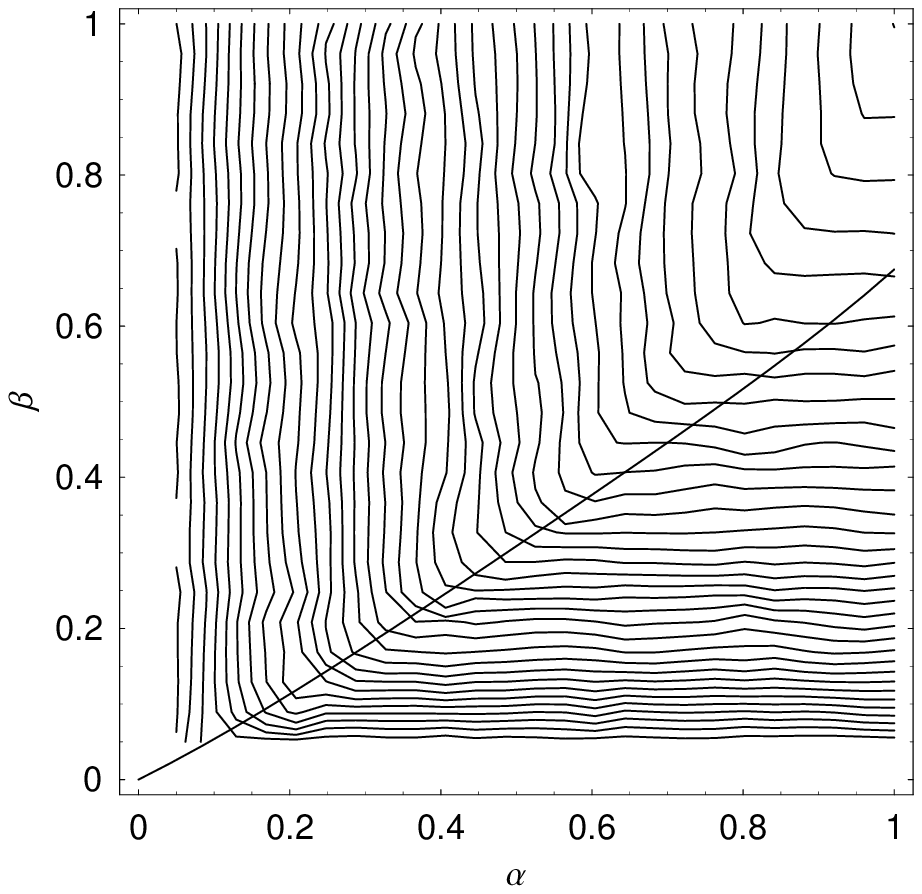}\\
(g)$q = 0.0$, $r = 1.0$\\
  \end{center}
 \end{minipage}
&
 \begin{minipage}{0.30\hsize}
  \begin{center}
   \includegraphics[width=\hsize]{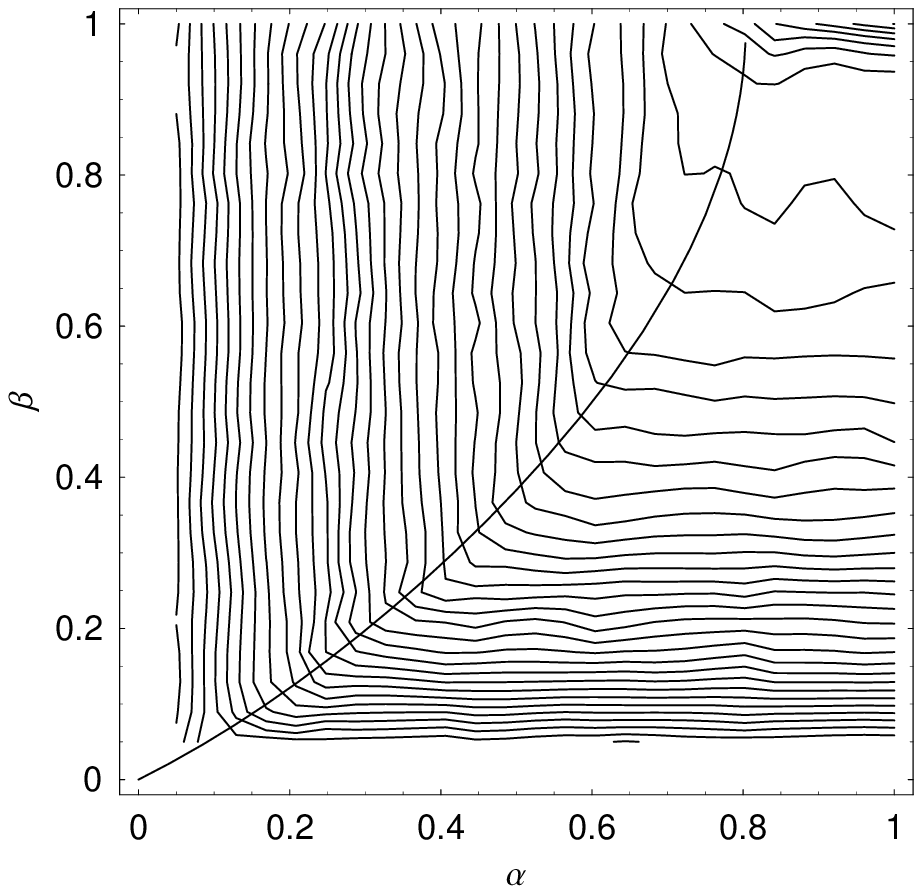}\\
(h)$q = 0.5$, $r = 1.0$\\
  \end{center}
 \end{minipage}
&
 \begin{minipage}{0.30\hsize}
  \begin{center}
   \includegraphics[width=\hsize]{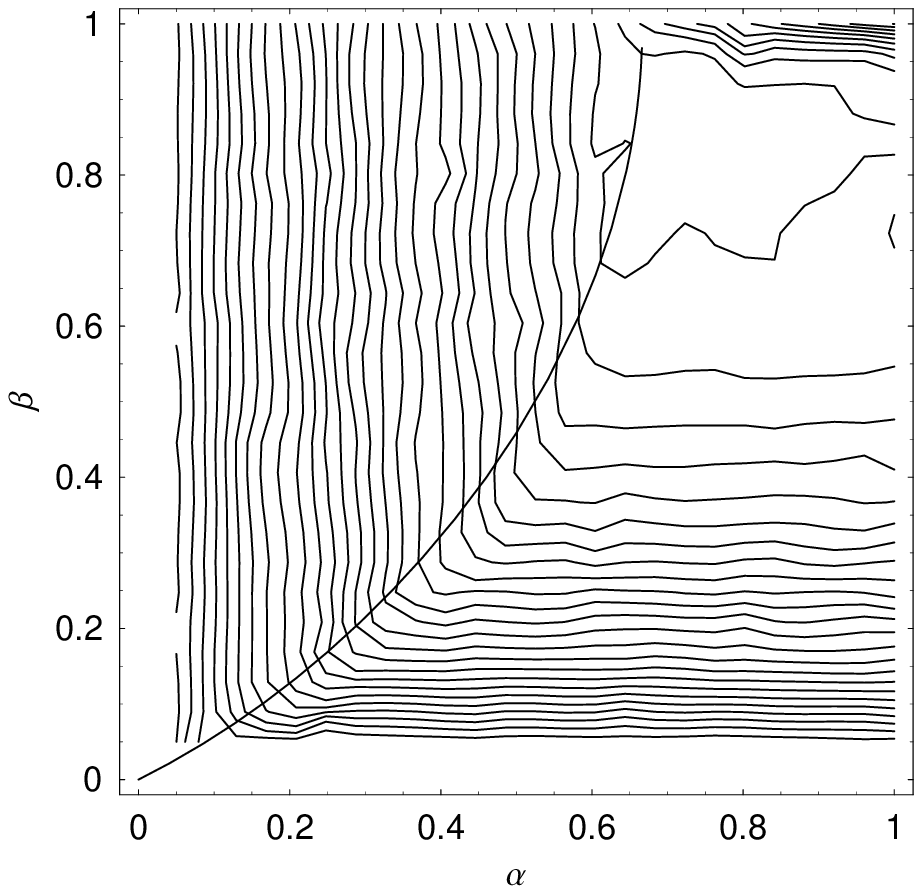}\\
(i)$q = 1.0$, $r = 1.0$\\
  \end{center}
 \end{minipage}
\end{tabular}
\end{center}
\caption{Approximate phase transition curves that are
calculated by \mbox{(\ref{eq:beta-alpha1})}, and
contour curves of
flow-$\alpha$-$\beta$ diagram \mbox{(Fig. \ref{fig:S-NFS_fab})}.
The theoretical curves are in good agreement with numerical results.}
\label{fig:S-NFS_PD}
\end{figure}

It should be noted that \mbox{(\ref{eq:rho_lr})}
and \mbox{(\ref{eq:beta-alpha1})} are a singular when $r = 0.0$.
Expression \mbox{(\ref{eq:beta-alpha1})} becomes
\begin{eqnarray}
\beta &=&
\frac{x\alpha}{x(1+\alpha)-\alpha}
\end{eqnarray}
in this singular case.
Figure \ref{fig:S-NFS_PD} gives a comparison between the approximate
phase transition curve \mbox{(\ref{eq:beta-alpha1})} and contour curves
of flow-$\alpha$-$\beta$ diagram obtained from numerical simulations.
The phase transition curves are very close to those
obtained from numerical simulations.
However we note that the shape of jamming line in the fundamental diagram is
assumed to be straight as \mbox{(\ref{eq:flow-rho})}.
Therefore we see the approximate result is not good in the
high flow region (where $\alpha$ and $\beta$ are close to $1$)
for figure \ref{fig:S-NFS_PD}-(d) because in this case the shape of
jamming line is apparently roundish (see figure \ref{fig:S-NFS_FD}-(d)).

The phase diagram consists of LD and HD phases,
but maximal-current (MC) phase does not appear.
Driver's anticipation effect make the area of HD phase smaller whereas
the slow-to-start effect gives the opposite tendency.
Phase transition curve which is straight for ASEP is bent downward
by driver's anticipation effect.
However, if $q \neq 0$, the slow-to-start effect becomes dominant
as flow increases.

The article \cite{Cecile} also discusses a phase diagram of a model
which includes only slow-to-start effect.
The structure of phase diagram is similar to those of
figure \ref{fig:S-NFS_PD}-(c).
S-NFS model shows more variety of phase diagrams.

\begin{figure}[thp]
\begin{center}
\includegraphics[width=0.5\linewidth]{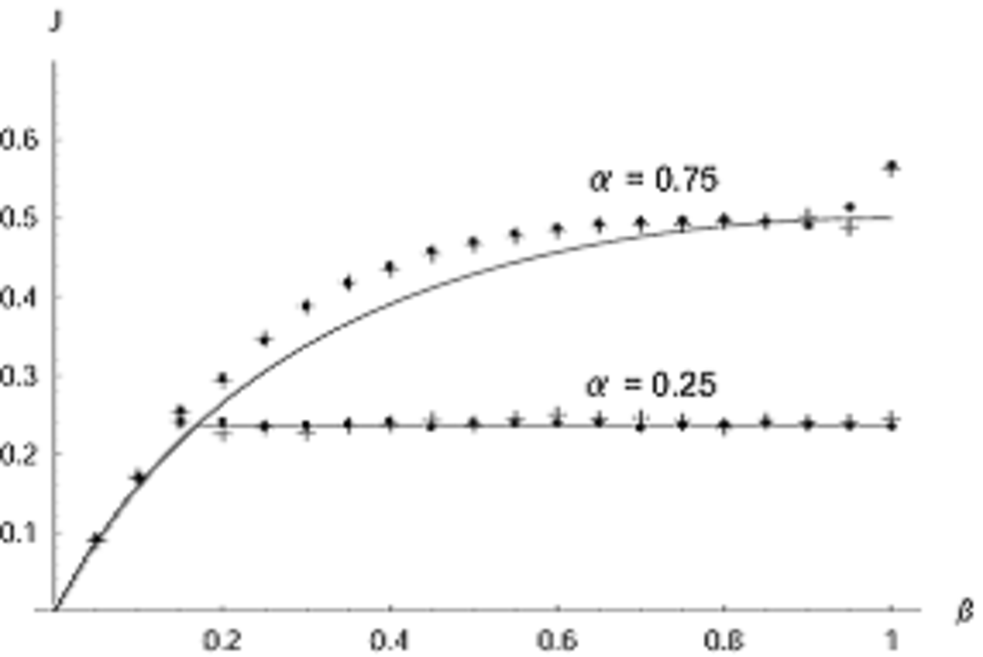}
\end{center}
\caption{Flow is plotted against $\beta$
at $\alpha = 0.25$ and $\alpha = 0.75$ when $p = q = r = 1.0$.
The symbols $+$ ($\bullet$) represent data for $L = 600$ ($L = 3000$). 
Solid lines denote the approximate results 
\mbox{(\ref{eq:flow-rho})} with \mbox{(\ref{eq:rho_lr})},
\mbox{(\ref{eq:rho_lr2})} and \mbox{(\ref{eq:beta-alpha1})}.}
\label{fig:S-NFS_PD_beta}
\end{figure}

According to \cite{Cecile}, for large $\beta$ there appears the enhancement
of the flow which looks like MC phase.
The enhancement is due to the finite system size effect
and vanishes with increasing
system size.
Figure \ref{fig:S-NFS_PD_beta} demonstrates the effects of system sizes on
our results.
The shape of flow does not change with increasing the system size.
Since there does not appear a plateau at $\alpha = 0.75$,
we judge that MC phase does not exist when $p = 1.0$ for S-NFS model.

\section{Summary and conclusions}
In this paper, we have considered a stochastic extension of
a traffic cellular automaton model recently proposed by
one of the authors \cite{NFS}.
This model (which we call S-NFS model) contains three parameters
which control random braking, slow-to-start, and driver's perspective
effects.
With special choice of these parameters, S-NFS model is reduced to previously
known traffic CA models such as NS model, QS model, SlS model etc..
Hence S-NFS model can be considered as an unified model of
these previously known models.
Next, we investigated fundamental diagrams of S-NFS model.
The shape of fundamental diagram of S-NFS model is similar to that with
use of empirical data.
Especially the metastable branches, which are indispensable to reproduce
empirical fundamental diagrams, clearly seen even when stochastic effects
are present.
This robustness of metastable branches in this model is advantageous
because empirical data shows metastability even though
stochastic effects always exist in real traffic.
Thus we can expect that S-NFS model captures essential feature of
empirical traffic flow and is very useful for investigating
various traffic phenomena such as jamming phase transition as well as for
application in traffic engineering.
Finally, we investigated phase diagrams of S-NFS model
with $V_{\max} = 1$ and $p = 1.0$.
The analytic expression of phase transition curve is obtained from
the approximate relations between flow and density near the boundary,
together with approximate gradient of jamming lines in fundamental diagrams.
The analytic phase transition curve successfully explains
those of numerical simulations.
Recently we have found that, in the case $p < 1.0$ MC phase appears
in the S-NFS model.
The analysis on the cases $V_{\max} > 1$ or $p < 1.0$ will be reported
elsewhere \cite{Sakai}.

\ack
The authors thank the Japan Highway Public Corporation
for providing us with the observed data. A part of this work is
financially supported by a Grand-in-Aid from the Ministry
of Education, Culture, Sports, Science and Technology, Japan
(No.18560053).

\end{document}

%% file: figure2.tex
\unitlength 0.1in
\begin{picture}( 44.0000, 38.0000)(  9.0000,-40.0000)
%
\special{pn 20}%
\special{sh 0.300}%
\special{pa 2100 1000}%
\special{pa 3100 1000}%
\special{pa 3100 1400}%
\special{pa 2100 1400}%
\special{pa 2100 1000}%
\special{fp}%
\put(26.0000,-12.0000){\makebox(0,0){S-NFS model}}%
\put(29.0000,-38.0000){\makebox(0,0){Rule-$184$}}%
\put(48.0000,-22.0000){\makebox(0,0){QS model}}%
\put(46.0000,-12.0000){\makebox(0,0){SlS model}}%
\put(14.0000,-22.0000){\makebox(0,0){NS model}}%
%
\special{pn 8}%
\special{pa 2400 3600}%
\special{pa 3400 3600}%
\special{pa 3400 4000}%
\special{pa 2400 4000}%
\special{pa 2400 3600}%
\special{fp}%
%
\special{pn 8}%
\special{pa 4300 2000}%
\special{pa 5300 2000}%
\special{pa 5300 2400}%
\special{pa 4300 2400}%
\special{pa 4300 2000}%
\special{fp}%
%
\special{pn 8}%
\special{pa 4100 1000}%
\special{pa 5100 1000}%
\special{pa 5100 1400}%
\special{pa 4100 1400}%
\special{pa 4100 1000}%
\special{fp}%
%
\special{pn 8}%
\special{pa 900 2000}%
\special{pa 1900 2000}%
\special{pa 1900 2400}%
\special{pa 900 2400}%
\special{pa 900 2000}%
\special{fp}%
%
\special{pn 8}%
\special{pa 2100 1200}%
\special{pa 1400 1200}%
\special{fp}%
%
\special{pn 8}%
\special{pa 1400 1200}%
\special{pa 1400 2000}%
\special{fp}%
\special{sh 1}%
\special{pa 1400 2000}%
\special{pa 1420 1934}%
\special{pa 1400 1948}%
\special{pa 1380 1934}%
\special{pa 1400 2000}%
\special{fp}%
\put(18.1000,-13.1000){\makebox(0,0){$q = 0.0$}}%
%
\special{pn 8}%
\special{pa 1400 3200}%
\special{pa 1400 3800}%
\special{fp}%
%
\special{pn 8}%
\special{pa 1400 3800}%
\special{pa 2400 3800}%
\special{fp}%
\special{sh 1}%
\special{pa 2400 3800}%
\special{pa 2334 3780}%
\special{pa 2348 3800}%
\special{pa 2334 3820}%
\special{pa 2400 3800}%
\special{fp}%
%
\special{pn 8}%
\special{pa 3100 1200}%
\special{pa 4100 1200}%
\special{fp}%
\special{sh 1}%
\special{pa 4100 1200}%
\special{pa 4034 1180}%
\special{pa 4048 1200}%
\special{pa 4034 1220}%
\special{pa 4100 1200}%
\special{fp}%
%
\put(36.0000,-13.0000){\makebox(0,0)[lb]{}}%
\put(36.0000,-13.0000){\makebox(0,0){$p = 1.0$}}%
\put(36.0000,-14.5000){\makebox(0,0){$q = 1.0$}}%
\put(14.5000,-27.0000){\makebox(0,0)[lb]{$V_{\max} = 1$}}%
\put(14.5000,-35.0000){\makebox(0,0)[lb]{$p = 1.0$}}%
%
\special{pn 8}%
\special{pa 3000 1400}%
\special{pa 3000 2200}%
\special{fp}%
%
\special{pn 8}%
\special{pa 3000 2200}%
\special{pa 4300 2200}%
\special{fp}%
\special{sh 1}%
\special{pa 4300 2200}%
\special{pa 4234 2180}%
\special{pa 4248 2200}%
\special{pa 4234 2220}%
\special{pa 4300 2200}%
\special{fp}%
\put(42.0000,-22.0000){\makebox(0,0)[rb]{$p = 1.0$}}%
\put(42.0000,-24.0000){\makebox(0,0)[rb]{$q = 0.0$}}%
\put(42.0000,-25.5000){\makebox(0,0)[rb]{$r = 1.0$}}%
%
\special{pn 8}%
\special{pa 2596 1400}%
\special{pa 2596 3600}%
\special{fp}%
\special{sh 1}%
\special{pa 2596 3600}%
\special{pa 2616 3534}%
\special{pa 2596 3548}%
\special{pa 2576 3534}%
\special{pa 2596 3600}%
\special{fp}%
\put(26.4500,-27.0000){\makebox(0,0)[lt]{$V_{\max} = 1$}}%
\put(26.5000,-29.0000){\makebox(0,0)[lt]{$p = 1.0$}}%
\put(26.5000,-31.0000){\makebox(0,0)[lt]{$q = r = 0.0$}}%
\put(26.0000,-4.0000){\makebox(0,0){NFS model}}%
%
\special{pn 8}%
\special{pa 2100 200}%
\special{pa 3100 200}%
\special{pa 3100 600}%
\special{pa 2100 600}%
\special{pa 2100 200}%
\special{fp}%
%
\special{pn 8}%
\special{pa 2600 1000}%
\special{pa 2600 600}%
\special{fp}%
\special{sh 1}%
\special{pa 2600 600}%
\special{pa 2580 668}%
\special{pa 2600 654}%
\special{pa 2620 668}%
\special{pa 2600 600}%
\special{fp}%
\put(26.5000,-7.0000){\makebox(0,0)[lt]{$p = q = r = 1.0$}}%
%
\special{pn 8}%
\special{pa 900 2800}%
\special{pa 1900 2800}%
\special{pa 1900 3200}%
\special{pa 900 3200}%
\special{pa 900 2800}%
\special{fp}%
\put(14.0000,-30.0000){\makebox(0,0){ASEP}}%
%
\special{pn 8}%
\special{pa 1400 2400}%
\special{pa 1400 2800}%
\special{fp}%
\special{sh 1}%
\special{pa 1400 2800}%
\special{pa 1420 2734}%
\special{pa 1400 2748}%
\special{pa 1380 2734}%
\special{pa 1400 2800}%
\special{fp}%
\put(36.0000,-16.0000){\makebox(0,0){$r = 0.0$}}%
\put(36.0000,-11.0000){\makebox(0,0){$V_{\max} = 1$}}%
\put(42.0000,-20.5000){\makebox(0,0)[rb]{$V_{\max} = 1$}}%
\put(18.1000,-14.6000){\makebox(0,0){$r = 0.0$}}%
%
\special{pn 8}%
\special{pa 900 400}%
\special{pa 1900 400}%
\special{pa 1900 800}%
\special{pa 900 800}%
\special{pa 900 400}%
\special{fp}%
%
\special{pn 8}%
\special{pa 1100 2000}%
\special{pa 1100 800}%
\special{fp}%
\special{sh 1}%
\special{pa 1100 800}%
\special{pa 1080 868}%
\special{pa 1100 854}%
\special{pa 1120 868}%
\special{pa 1100 800}%
\special{fp}%
\put(11.5000,-9.0000){\makebox(0,0)[lt]{$p = 1.0$}}%
\put(14.0000,-6.0000){\makebox(0,0){mFI model}}%
\end{picture}%

%% file: figure5.tex
\unitlength 0.1in
\begin{picture}( 44.0600, 12.5000)(  4.6200,-16.5000)
%
\special{pn 13}%
\special{pa 516 1120}%
\special{pa 1034 1120}%
\special{pa 1034 1380}%
\special{pa 516 1380}%
\special{pa 516 1120}%
\special{fp}%
%
\special{pn 8}%
\special{pa 776 1120}%
\special{pa 776 1380}%
\special{fp}%
%
\special{pn 8}%
\special{ar 906 1250 110 110  0.0000000 6.2831853}%
%
\special{pn 8}%
\special{ar 646 1250 110 110  0.0000000 6.2831853}%
%
\special{pn 13}%
\special{pa 3738 1120}%
\special{pa 4256 1120}%
\special{pa 4256 1380}%
\special{pa 3738 1380}%
\special{pa 3738 1120}%
\special{fp}%
%
\special{pn 8}%
\special{ar 3868 1250 110 110  0.0000000 6.2831853}%
%
\special{pn 8}%
\special{ar 4128 1250 110 110  0.0000000 6.2831853}%
%
\special{pn 8}%
\special{pa 3998 1120}%
\special{pa 3998 1380}%
\special{fp}%
%
\special{pn 13}%
\special{pa 4350 1120}%
\special{pa 4868 1120}%
\special{pa 4868 1380}%
\special{pa 4350 1380}%
\special{pa 4350 1120}%
\special{fp}%
%
\special{pn 8}%
\special{sh 0.600}%
\special{ar 4480 1250 110 110  0.0000000 6.2831853}%
%
\special{pn 8}%
\special{sh 0.600}%
\special{ar 4740 1250 110 110  0.0000000 6.2831853}%
%
\special{pn 8}%
\special{pa 4610 1120}%
\special{pa 4610 1380}%
\special{fp}%
%
\special{pn 13}%
\special{pa 1128 1120}%
\special{pa 3648 1120}%
\special{pa 3648 1480}%
\special{pa 1128 1480}%
\special{pa 1128 1120}%
\special{fp}%
%
\special{pn 8}%
\special{pa 1488 1120}%
\special{pa 1488 1480}%
\special{fp}%
\special{pa 1848 1480}%
\special{pa 1848 1120}%
\special{fp}%
\special{pa 2208 1120}%
\special{pa 2208 1480}%
\special{fp}%
\special{pa 3288 1480}%
\special{pa 3288 1120}%
\special{fp}%
%
\special{pn 8}%
\special{sh 0.600}%
\special{ar 2028 1300 154 154  0.0000000 6.2831853}%
\put(6.4200,-10.3000){\makebox(0,0){$-2$}}%
\put(13.0800,-10.3000){\makebox(0,0){$0$}}%
\put(34.6800,-10.3000){\makebox(0,0){$L\!-\!1$}}%
\put(41.2500,-10.3000){\makebox(0,0){$L\!\!+\!\!1$}}%
\put(47.3700,-10.3000){\makebox(0,0){$L\!\!+\!\!3$}}%
%
\special{pn 13}%
\special{pa 516 1120}%
\special{pa 516 400}%
\special{da 0.070}%
\special{pa 4260 400}%
\special{pa 4260 1120}%
\special{da 0.070}%
%
\special{pn 13}%
\special{pa 1128 670}%
\special{pa 1128 1120}%
\special{da 0.070}%
\special{pa 3648 1120}%
\special{pa 3648 670}%
\special{da 0.070}%
%
\special{pn 8}%
\special{pa 1470 490}%
\special{pa 570 490}%
\special{fp}%
\special{sh 1}%
\special{pa 570 490}%
\special{pa 638 510}%
\special{pa 624 490}%
\special{pa 638 470}%
\special{pa 570 490}%
\special{fp}%
\special{pa 1470 490}%
\special{pa 4206 490}%
\special{fp}%
\special{sh 1}%
\special{pa 4206 490}%
\special{pa 4140 470}%
\special{pa 4154 490}%
\special{pa 4140 510}%
\special{pa 4206 490}%
\special{fp}%
%
\special{pn 8}%
\special{pa 1668 760}%
\special{pa 3594 760}%
\special{fp}%
\special{sh 1}%
\special{pa 3594 760}%
\special{pa 3528 740}%
\special{pa 3542 760}%
\special{pa 3528 780}%
\special{pa 3594 760}%
\special{fp}%
\special{pa 1668 760}%
\special{pa 1174 760}%
\special{fp}%
\special{sh 1}%
\special{pa 1174 760}%
\special{pa 1240 780}%
\special{pa 1226 760}%
\special{pa 1240 740}%
\special{pa 1174 760}%
\special{fp}%
\put(23.8800,-5.8000){\makebox(0,0){\small (a)}}%
\put(23.8800,-8.5000){\makebox(0,0){\small (b)}}%
%
\special{pn 8}%
\special{pa 2030 1300}%
\special{pa 2430 1300}%
\special{fp}%
\special{sh 1}%
\special{pa 2430 1300}%
\special{pa 2364 1280}%
\special{pa 2378 1300}%
\special{pa 2364 1320}%
\special{pa 2430 1300}%
\special{fp}%
%
\special{pn 8}%
\special{pa 650 1600}%
\special{pa 650 1250}%
\special{fp}%
\special{sh 1}%
\special{pa 650 1250}%
\special{pa 630 1318}%
\special{pa 650 1304}%
\special{pa 670 1318}%
\special{pa 650 1250}%
\special{fp}%
%
\special{pn 8}%
\special{pa 910 1600}%
\special{pa 910 1250}%
\special{fp}%
\special{sh 1}%
\special{pa 910 1250}%
\special{pa 890 1318}%
\special{pa 910 1304}%
\special{pa 930 1318}%
\special{pa 910 1250}%
\special{fp}%
%
\special{pn 8}%
\special{pa 3870 1600}%
\special{pa 3870 1250}%
\special{fp}%
\special{sh 1}%
\special{pa 3870 1250}%
\special{pa 3850 1318}%
\special{pa 3870 1304}%
\special{pa 3890 1318}%
\special{pa 3870 1250}%
\special{fp}%
%
\special{pn 8}%
\special{pa 4130 1600}%
\special{pa 4130 1250}%
\special{fp}%
\special{sh 1}%
\special{pa 4130 1250}%
\special{pa 4110 1318}%
\special{pa 4130 1304}%
\special{pa 4150 1318}%
\special{pa 4130 1250}%
\special{fp}%
\put(6.0000,-16.5000){\makebox(0,0)[lt]{$\alpha$}}%
\put(8.6000,-16.5000){\makebox(0,0)[lt]{$\alpha$}}%
\put(39.7000,-16.5000){\makebox(0,0)[rt]{$1\!\!-\!\!\beta$}}%
\put(42.8000,-16.5000){\makebox(0,0)[rt]{$1\!\!-\!\!\beta$}}%
\end{picture}%

%% file: figure8a.tex
\unitlength 0.1in
\begin{picture}( 10.9900,  5.7000)(  2.7600, -7.5000)
%
\special{pn 20}%
\special{pa 326 330}%
\special{pa 588 330}%
\special{pa 588 592}%
\special{pa 326 592}%
\special{pa 326 330}%
\special{fp}%
%
\special{pn 20}%
\special{pa 720 592}%
\special{pa 720 330}%
\special{fp}%
\special{pa 720 330}%
\special{pa 1376 330}%
\special{fp}%
\special{pa 1376 592}%
\special{pa 720 592}%
\special{fp}%
%
\special{pn 8}%
\special{pa 982 330}%
\special{pa 982 592}%
\special{fp}%
\put(4.5600,-2.6500){\makebox(0,0){$-1$}}%
\put(8.4900,-2.6500){\makebox(0,0){$0$}}%
%
\special{pn 8}%
\special{ar 456 462 112 112  0.0000000 6.2831853}%
%
\special{pn 8}%
\special{pa 456 724}%
\special{pa 456 462}%
\special{fp}%
\special{sh 1}%
\special{pa 456 462}%
\special{pa 436 528}%
\special{pa 456 514}%
\special{pa 476 528}%
\special{pa 456 462}%
\special{fp}%
%
\special{pn 13}%
\special{pa 654 200}%
\special{pa 654 724}%
\special{da 0.070}%
%
\special{pn 8}%
\special{pa 522 526}%
\special{pa 786 724}%
\special{dt 0.045}%
\special{sh 1}%
\special{pa 786 724}%
\special{pa 744 668}%
\special{pa 742 692}%
\special{pa 720 700}%
\special{pa 786 724}%
\special{fp}%
\put(4.0000,-7.5000){\makebox(0,0)[lt]{$\alpha$}}%
\end{picture}%

%% file: figure8b.tex
\unitlength 0.1in
\begin{picture}( 10.9900,  5.7000)(  2.7600, -7.5000)
%
\special{pn 20}%
\special{pa 326 330}%
\special{pa 588 330}%
\special{pa 588 592}%
\special{pa 326 592}%
\special{pa 326 330}%
\special{fp}%
%
\special{pn 20}%
\special{pa 720 592}%
\special{pa 720 330}%
\special{fp}%
\special{pa 720 330}%
\special{pa 1376 330}%
\special{fp}%
\special{pa 1376 592}%
\special{pa 720 592}%
\special{fp}%
%
\special{pn 8}%
\special{pa 1244 330}%
\special{pa 1244 592}%
\special{fp}%
\put(4.5600,-2.6500){\makebox(0,0){$-1$}}%
\put(8.4900,-2.6500){\makebox(0,0){$0$}}%
%
\special{pn 8}%
\special{ar 456 462 112 112  0.0000000 6.2831853}%
%
\special{pn 8}%
\special{pa 456 724}%
\special{pa 456 462}%
\special{fp}%
\special{sh 1}%
\special{pa 456 462}%
\special{pa 436 528}%
\special{pa 456 514}%
\special{pa 476 528}%
\special{pa 456 462}%
\special{fp}%
\put(4.0000,-7.5000){\makebox(0,0)[lt]{$\alpha$}}%
%
\special{pn 8}%
\special{sh 0.600}%
\special{ar 850 462 112 112  0.0000000 6.2831853}%
%
\special{pn 13}%
\special{pa 654 200}%
\special{pa 654 724}%
\special{da 0.070}%
%
\special{pn 8}%
\special{pa 522 526}%
\special{pa 786 724}%
\special{dt 0.045}%
\special{sh 1}%
\special{pa 786 724}%
\special{pa 744 668}%
\special{pa 742 692}%
\special{pa 720 700}%
\special{pa 786 724}%
\special{fp}%
%
\special{pn 8}%
\special{pa 916 526}%
\special{pa 1178 724}%
\special{dt 0.045}%
\special{sh 1}%
\special{pa 1178 724}%
\special{pa 1138 668}%
\special{pa 1136 692}%
\special{pa 1114 700}%
\special{pa 1178 724}%
\special{fp}%
%
\special{pn 20}%
\special{pa 326 330}%
\special{pa 588 330}%
\special{pa 588 592}%
\special{pa 326 592}%
\special{pa 326 330}%
\special{fp}%
%
\special{pn 20}%
\special{pa 720 592}%
\special{pa 720 330}%
\special{fp}%
\special{pa 720 330}%
\special{pa 1376 330}%
\special{fp}%
\special{pa 1376 592}%
\special{pa 720 592}%
\special{fp}%
%
\special{pn 8}%
\special{pa 982 330}%
\special{pa 982 592}%
\special{fp}%
\put(4.5600,-2.6500){\makebox(0,0){$-1$}}%
\put(8.4900,-2.6500){\makebox(0,0){$0$}}%
%
\special{pn 8}%
\special{ar 456 462 112 112  0.0000000 6.2831853}%
%
\special{pn 8}%
\special{pa 456 724}%
\special{pa 456 462}%
\special{fp}%
\special{sh 1}%
\special{pa 456 462}%
\special{pa 436 528}%
\special{pa 456 514}%
\special{pa 476 528}%
\special{pa 456 462}%
\special{fp}%
\end{picture}%

%% file: figure8c.tex
\unitlength 0.1in
\begin{picture}( 10.4900,  5.5000)(  4.0000, -7.3000)
%
\special{pn 20}%
\special{pa 1188 334}%
\special{pa 1450 334}%
\special{pa 1450 598}%
\special{pa 1188 598}%
\special{pa 1188 334}%
\special{fp}%
%
\special{pn 20}%
\special{pa 1056 330}%
\special{pa 1056 592}%
\special{fp}%
\special{pa 1056 592}%
\special{pa 400 592}%
\special{fp}%
\special{pa 400 330}%
\special{pa 1056 330}%
\special{fp}%
%
\special{pn 8}%
\special{pa 794 330}%
\special{pa 794 592}%
\special{fp}%
\put(13.1900,-2.6800){\makebox(0,0){$L$}}%
\put(9.2500,-2.6500){\makebox(0,0){$L\!-\!1$}}%
%
\special{pn 8}%
\special{sh 0.600}%
\special{ar 926 462 112 112  0.0000000 6.2831853}%
%
\special{pn 8}%
\special{pa 1320 724}%
\special{pa 1320 462}%
\special{fp}%
\special{sh 1}%
\special{pa 1320 462}%
\special{pa 1300 528}%
\special{pa 1320 514}%
\special{pa 1340 528}%
\special{pa 1320 462}%
\special{fp}%
\put(13.2000,-8.0000){\makebox(0,0){$1\!\!-\!\!\beta$}}%
%
\special{pn 13}%
\special{pa 1122 200}%
\special{pa 1122 724}%
\special{da 0.070}%
%
\special{pn 8}%
\special{pa 990 534}%
\special{pa 1254 730}%
\special{dt 0.045}%
\special{sh 1}%
\special{pa 1254 730}%
\special{pa 1212 674}%
\special{pa 1210 698}%
\special{pa 1188 706}%
\special{pa 1254 730}%
\special{fp}%
\end{picture}%

%% file: figure8d.tex
\unitlength 0.1in
\begin{picture}( 13.1200,  5.5000)(  4.0000, -7.3000)
%
\special{pn 20}%
\special{pa 1188 334}%
\special{pa 1712 334}%
\special{pa 1712 598}%
\special{pa 1188 598}%
\special{pa 1188 334}%
\special{fp}%
%
\special{pn 20}%
\special{pa 1056 330}%
\special{pa 1056 592}%
\special{fp}%
\special{pa 1056 592}%
\special{pa 400 592}%
\special{fp}%
\special{pa 400 330}%
\special{pa 1056 330}%
\special{fp}%
%
\special{pn 8}%
\special{pa 794 330}%
\special{pa 794 592}%
\special{fp}%
\put(13.1900,-2.6800){\makebox(0,0){$L$}}%
\put(9.2500,-2.6500){\makebox(0,0){$L\!-\!1$}}%
%
\special{pn 8}%
\special{sh 0.600}%
\special{ar 926 462 112 112  0.0000000 6.2831853}%
%
\special{pn 8}%
\special{pa 1320 724}%
\special{pa 1320 462}%
\special{fp}%
\special{sh 1}%
\special{pa 1320 462}%
\special{pa 1300 528}%
\special{pa 1320 514}%
\special{pa 1340 528}%
\special{pa 1320 462}%
\special{fp}%
\put(13.2000,-8.0000){\makebox(0,0){$1\!\!-\!\!\beta$}}%
%
\special{pn 13}%
\special{pa 1122 200}%
\special{pa 1122 724}%
\special{da 0.070}%
%
\special{pn 8}%
\special{pa 990 534}%
\special{pa 1254 730}%
\special{dt 0.045}%
\special{sh 1}%
\special{pa 1254 730}%
\special{pa 1212 674}%
\special{pa 1210 698}%
\special{pa 1188 706}%
\special{pa 1254 730}%
\special{fp}%
%
\special{pn 8}%
\special{ar 1320 462 112 112  0.0000000 6.2831853}%
%
\special{pn 8}%
\special{pa 1450 330}%
\special{pa 1450 592}%
\special{fp}%
%
\special{pn 8}%
\special{pa 1582 724}%
\special{pa 1582 462}%
\special{fp}%
\special{sh 1}%
\special{pa 1582 462}%
\special{pa 1562 528}%
\special{pa 1582 514}%
\special{pa 1602 528}%
\special{pa 1582 462}%
\special{fp}%
\put(15.8000,-8.0000){\makebox(0,0){$1\!\!-\!\!\beta$}}%
\put(15.8100,-2.7100){\makebox(0,0){$L\!+\!1$}}%
%
\special{pn 8}%
\special{pa 1386 534}%
\special{pa 1648 730}%
\special{dt 0.045}%
\special{sh 1}%
\special{pa 1648 730}%
\special{pa 1606 674}%
\special{pa 1604 698}%
\special{pa 1582 706}%
\special{pa 1648 730}%
\special{fp}%
\end{picture}%

%% file: figure9a.tex
\unitlength 0.1in
\begin{picture}( 15.7000,  5.6000)(  3.7000, -7.6000)
%
\special{pn 8}%
\special{pa 540 200}%
\special{pa 1660 200}%
\special{fp}%
\special{pa 540 480}%
\special{pa 1940 480}%
\special{fp}%
\special{pa 1940 760}%
\special{pa 540 760}%
\special{fp}%
\special{pa 680 200}%
\special{pa 680 760}%
\special{fp}%
\special{pa 960 760}%
\special{pa 960 200}%
\special{fp}%
\special{pa 1240 200}%
\special{pa 1240 760}%
\special{fp}%
\special{pa 1520 760}%
\special{pa 1520 200}%
\special{fp}%
\special{pa 1800 480}%
\special{pa 1800 760}%
\special{fp}%
\put(4.7000,-5.8500){\makebox(0,0)[lt]{$t$}}%
\put(3.7000,-3.1000){\makebox(0,0)[lt]{$t\!-\!1$}}%
%
\special{pn 8}%
\special{sh 0.600}%
\special{ar 1380 620 106 106  0.0000000 6.2831853}%
%
\special{pn 8}%
\special{sh 0.300}%
\special{ar 1100 620 106 106  0.0000000 6.2831853}%
%
\special{pn 8}%
\special{sh 0.300}%
\special{ar 820 620 106 106  0.0000000 6.2831853}%
%
\special{pn 8}%
\special{sh 0.600}%
\special{ar 1380 340 106 106  0.0000000 6.2831853}%
%
\special{pn 8}%
\special{sh 0.300}%
\special{ar 1100 340 106 106  0.0000000 6.2831853}%
%
\special{pn 8}%
\special{sh 0.300}%
\special{ar 820 340 106 106  0.0000000 6.2831853}%
\end{picture}%

%% file: figure9b.tex
\unitlength 0.1in
\begin{picture}( 15.7000,  5.6000)(  3.7000, -7.6000)
%
\special{pn 8}%
\special{pa 540 200}%
\special{pa 1660 200}%
\special{fp}%
\special{pa 540 480}%
\special{pa 1940 480}%
\special{fp}%
\special{pa 1940 760}%
\special{pa 540 760}%
\special{fp}%
\special{pa 680 200}%
\special{pa 680 760}%
\special{fp}%
\special{pa 960 760}%
\special{pa 960 200}%
\special{fp}%
\special{pa 1240 200}%
\special{pa 1240 760}%
\special{fp}%
\special{pa 1520 760}%
\special{pa 1520 200}%
\special{fp}%
\special{pa 1800 480}%
\special{pa 1800 760}%
\special{fp}%
\put(4.7000,-5.8500){\makebox(0,0)[lt]{$t$}}%
\put(3.7000,-3.1000){\makebox(0,0)[lt]{$t\!-\!1$}}%
%
\special{pn 8}%
\special{sh 0.300}%
\special{ar 1660 620 106 106  0.0000000 6.2831853}%
%
\special{pn 8}%
\special{sh 0.600}%
\special{ar 1100 620 106 106  0.0000000 6.2831853}%
%
\special{pn 8}%
\special{sh 0.300}%
\special{ar 820 620 106 106  0.0000000 6.2831853}%
%
\special{pn 8}%
\special{sh 0.600}%
\special{ar 1380 340 106 106  0.0000000 6.2831853}%
%
\special{pn 8}%
\special{sh 0.300}%
\special{ar 1100 340 106 106  0.0000000 6.2831853}%
%
\special{pn 8}%
\special{sh 0.300}%
\special{ar 820 340 106 106  0.0000000 6.2831853}%
\end{picture}%

%% file: figure9c.tex
\unitlength 0.1in
\begin{picture}( 15.7000,  5.6000)(  3.7000, -7.6000)
%
\special{pn 8}%
\special{pa 540 200}%
\special{pa 1660 200}%
\special{fp}%
\special{pa 540 480}%
\special{pa 1940 480}%
\special{fp}%
\special{pa 1940 760}%
\special{pa 540 760}%
\special{fp}%
\special{pa 680 200}%
\special{pa 680 760}%
\special{fp}%
\special{pa 960 760}%
\special{pa 960 200}%
\special{fp}%
\special{pa 1240 200}%
\special{pa 1240 760}%
\special{fp}%
\special{pa 1520 760}%
\special{pa 1520 200}%
\special{fp}%
\special{pa 1800 480}%
\special{pa 1800 760}%
\special{fp}%
\put(4.7000,-5.8500){\makebox(0,0)[lt]{$t$}}%
\put(3.7000,-3.1000){\makebox(0,0)[lt]{$t\!-\!1$}}%
%
\special{pn 8}%
\special{sh 0.300}%
\special{ar 1660 620 106 106  0.0000000 6.2831853}%
%
\special{pn 8}%
\special{sh 0.300}%
\special{ar 1380 620 106 106  0.0000000 6.2831853}%
%
\special{pn 8}%
\special{sh 0.600}%
\special{ar 820 620 106 106  0.0000000 6.2831853}%
%
\special{pn 8}%
\special{sh 0.600}%
\special{ar 1380 340 106 106  0.0000000 6.2831853}%
%
\special{pn 8}%
\special{sh 0.300}%
\special{ar 1100 340 106 106  0.0000000 6.2831853}%
%
\special{pn 8}%
\special{sh 0.300}%
\special{ar 820 340 106 106  0.0000000 6.2831853}%
\end{picture}%

%% file: figure10.tex
\unitlength 0.1in
\begin{picture}( 26.7400, 10.4000)(  0.8500,-11.9000)
%
\special{pn 20}%
\special{pa 520 190}%
\special{pa 2760 190}%
\special{fp}%
\put(20.2300,-9.9000){\makebox(0,0)[lt]{$u_1(t+1)$}}%
\put(26.6300,-3.1000){\makebox(0,0)[lt]{$u_2(t)$}}%
%
\special{pn 20}%
\special{pa 640 470}%
\special{pa 2000 470}%
\special{pa 2000 1190}%
\special{pa 640 1190}%
\special{pa 640 470}%
\special{da 0.070}%
%
\special{pn 20}%
\special{pa 1280 150}%
\special{pa 2640 150}%
\special{pa 2640 870}%
\special{pa 1280 870}%
\special{pa 1280 150}%
\special{dt 0.054}%
\put(4.9000,-9.9000){\makebox(0,0){$t\!+\!1$}}%
\put(4.9000,-6.7000){\makebox(0,0){$t$}}%
\put(4.9000,-3.5000){\makebox(0,0){$t\!-\!1$}}%
\put(11.5900,-9.9000){\makebox(0,0){B}}%
\put(17.9900,-9.9000){\makebox(0,0){A}}%
\put(11.5900,-6.7000){\makebox(0,0){B}}%
\put(14.7900,-6.7000){\makebox(0,0){A}}%
%
\special{pn 8}%
\special{ar 1800 990 120 120  0.0000000 6.2831853}%
%
\special{pn 8}%
\special{ar 1160 990 120 120  0.0000000 6.2831853}%
%
\special{pn 8}%
\special{ar 840 990 120 120  0.0000000 6.2831853}%
%
\special{pn 8}%
\special{ar 840 670 120 120  0.0000000 6.2831853}%
%
\special{pn 8}%
\special{ar 1160 670 120 120  0.0000000 6.2831853}%
%
\special{pn 8}%
\special{ar 1480 670 120 120  0.0000000 6.2831853}%
%
\special{pn 8}%
\special{ar 2120 670 120 120  0.0000000 6.2831853}%
%
\special{pn 8}%
\special{ar 2440 670 120 120  0.0000000 6.2831853}%
%
\special{pn 8}%
\special{ar 2120 350 120 120  0.0000000 6.2831853}%
%
\special{pn 8}%
\special{ar 1800 350 120 120  0.0000000 6.2831853}%
%
\special{pn 8}%
\special{ar 1480 350 120 120  0.0000000 6.2831853}%
%
\special{pn 8}%
\special{ar 1160 350 120 120  0.0000000 6.2831853}%
%
\special{pn 8}%
\special{ar 840 350 120 120  0.0000000 6.2831853}%
%
\special{pn 20}%
\special{pa 2760 510}%
\special{pa 520 510}%
\special{fp}%
%
\special{pn 20}%
\special{pa 520 830}%
\special{pa 2760 830}%
\special{fp}%
%
\special{pn 20}%
\special{pa 2120 1150}%
\special{pa 520 1150}%
\special{fp}%
%
\special{pn 8}%
\special{pa 680 1150}%
\special{pa 680 190}%
\special{fp}%
%
\special{pn 8}%
\special{pa 1000 190}%
\special{pa 1000 1150}%
\special{fp}%
%
\special{pn 8}%
\special{pa 1320 1150}%
\special{pa 1320 190}%
\special{fp}%
%
\special{pn 8}%
\special{pa 1640 190}%
\special{pa 1640 1150}%
\special{fp}%
%
\special{pn 8}%
\special{pa 1960 1150}%
\special{pa 1960 190}%
\special{fp}%
%
\special{pn 8}%
\special{pa 2280 190}%
\special{pa 2280 830}%
\special{fp}%
%
\special{pn 8}%
\special{pa 2600 830}%
\special{pa 2600 190}%
\special{fp}%
\end{picture}%